# Evidence for Widespread Hydrogen Sequestration within the Moon's South Pole Cold Traps

T. P. McClanahan[1,6], A. M. Parsons[1], T. A. Livengood[1,2], J. J. Su[3], G. Chin[1], D. Hamara[4], K. Harshman[4], and R. D. Starr[1,5]
[1] Solar System Exploration Division, NASA Goddard Space Flight Center, Greenbelt, MD 20771, USA; timothy.p.mcclanahan@nasa.gov
[2] University of Maryland, College Park, MD 20742, USA
[3] Systems Engineering Group Inc., Columbia, MD 21046, USA
[4] Lunar and Planetary Laboratory, University of Arizona, Tucson, AZ 85719, USA
[5] Catholic University of America, Washington, DC 20064, USA

## Abstract

Hydrogen-bearing volatiles are observed to be concentrated, likely in the form of water ice, within most of the Moon's permanently shadowed regions (PSRs), poleward of 77° S. Results show that instrumental blurring of the Moon's epithermal neutron flux correlates the PSRs' observed hydrogen concentration by their areal density. Epithermal neutron observations of 502 PSRs are positively correlated indicating that they have similar expected hydrogen concentrations, $0.28 \pm 0.03$ wt% water-equivalent hydrogen, relative to neutron background observations (lower bounds). The correlation arises from the PSRs' proportional detection attributed to their similar hydrogen distributions and their areal fraction of the collimated instrument footprint of the Collimated Sensor for Epithermal Neutrons (CSETN), which is part of the Lunar Exploration Neutron Detector on board the Lunar Reconnaissance Orbiter (LRO). The lowest hydrogen concentration areas coincide with low PSR areal densities that occur with highly illuminated and warm, equator-facing sloped surfaces. Results show that the maximum hydrogen concentrations observed within the Haworth, Shoemaker, and Faustini PSRs coincide with their coldest surface temperatures, below 75 K that occur near the base of their poleward-facing slopes. Anomalously enhanced hydrogen concentrations around the Cabeus-1 PSR suggest at least two lunar hydrogen sources. The uncollimated neutron counting rate map is subtracted from CSETN's collimated neutron map using a novel spatial bandpass filter. The results indicate water ice and perhaps other hydrogen-bearing volatiles are being randomly distributed to the surface and the PSRs' low sublimation rates likely maximize their residence times and elevate their surface concentrations. CSETN's corrected south polar map is correlated to coregistered maximum temperature and topography maps made by LRO's Diviner and Lunar Orbiter Laser Altimeter instruments.

*Unified Astronomy Thesaurus concepts:* Lunar surface (974); Lunar origin (966); Lunar composition (948); Ice formation (2092)

## 1. Introduction

Over the last few decades, several independent studies have found evidence that heterogeneous distributions of water ice and other hydrogen-bearing volatiles exist toward the Moon's poles (Feldman et al. 1998; Clark 2009; Pieters et al. 2009; Sunshine et al. 2009; Colaprete et al. 2010; Gladstone et al. 2010; Mitrofanov et al. 2010b; Hayne et al. 2015; Li & Milliken 2017; Li et al. 2018). These results have been used to hone NASA's future planetary mission objectives in its ongoing quest to return humans to the lunar surface and to understand the origins and processes that influence water distributions within the solar system. Finding efficiently extractable resources, in particular water, is a high-priority objective because their availability will determine the feasibility and longevity of the next generation of crewed missions to the lunar surface and beyond, as targeted by the present planetary decadal survey Origins, Worlds and Life: A Decadal Strategy for Planetary Science and Astrobiology (National Academies of Sciences, Engineering, and Medicine 2022), as well as the Artemis Science Definition Team: Mission Planning Update (Lawrence et al. 2023). Hydrogen-bearing volatile distributions are also of scientific interest because they may preserve a record of the 4.53 billion years of the Earth–Moon water budget, as well as provide clues to early and ongoing inner solar system formation processes (Zellar et al. 1966; Starukhina 2006; Barnes et al. 2016; Alexander 2017; Russell et al. 2017).

This paper investigates the spatial distribution of the Moon's south pole hydrogen-bearing volatiles. Our results show evidence that most permanently shadowed regions (PSRs) poleward of 77° S have enhanced hydrogen concentrations relative to that expected from surrounding non-PSR surfaces. The findings are consistent with the theoretical results of Watson et al. (1961), who predicted that water ice could be exclusively accumulated within PSRs.

Our hypothesis and model emulate the Watson et al. (1961) theoretical studies to predict how an areal distribution of similarly hydrogen-enhanced PSRs should be observed using epithermal neutrons. Results show that the PSRs and their internal cold traps maintain the highest concentration deposits of enhanced water-ice. Our observations show that instrumental blurring correlates the PSRs' hydrogen concentrations as a function of their areal density. Several independent lines of evidence demonstrate (1) a comprehensive new understanding of the Moon's south pole epithermal neutron-emission flux as characterized by surface illumination, topography, slope, and temperature, and (2) widespread evidence that hydrogen concentrations are similarly enhanced within the PSR, poleward of 77° S. The correlation is induced by the ratio of hydrogen-enhanced PSR area to anhydrous non-PSR area, observed within the collimated footprint area of the Collimated Sensor for Epithermal Neutrons (CSETN). CSETN is an instrument suite of the Lunar Exploration Neutron Detector (LEND), which operates on board the Lunar Reconnaissance Orbiter (LRO; Chin et al. 2007; Mitrofanov et al. 2010a; Paige et al. 2010a; Smith et al. 2010; Zuber et al. 2010).

---
[6] Code 690.1, Rm# E108, B34, NASA Goddard Space Flight Center, Greenbelt MD.

### 1.1. Background

The Moons' PSRs are found toward polar latitudes, and are coincident with surfaces that receive no direct solar irradiation. PSRs are formed by the combination of polar topographic depressions and the regions constantly low solar incidence angles attributed to the slight $1°.54$ tilt of the Moon's spin axis relative to the ecliptic plane. Watson et al. (1961) proposed from the Moon's vacuum and expected surface temperature distributions that highly nonlinear sublimation rates should occur for PSR and non-PSR surfaces. PSR maintain low maximum temperatures, < 120 K and may vary diurnally by a few K dependent on reflected light distributions. At such cryogenic temperatures water ice and hydrogen volatiles low sublimation rates yield high surface residence times that may approach billions of years. Alternatively, non-PSR surfaces are expected to be relatively anhydrous because their periodic illumination yields higher, insolation-dependent maximum temperatures, averaging 243 K poleward of 80 S by Diviner. The conditions indicate the non-PSR surfaces should have significantly higher sublimation rates than PSR, and an expectation of very short surface residence times that would preclude any significant accumulations.

PSR areas may range from the smallest, made by shadowing individual regolith grains on the surface, to an upper diameter limit of ∼37 km, at the Shoemaker crater's PSR. PSR diameters are derived for this study using "region growing" software that integrates PSR areas from contiguous PSR pixels, as derived from a binary PSR map. Pixels of 0% illumination are mapped from an averaged surface illumination map (Sonka et al. 1998; Mazarico et al. 2011). Cold-trap areas are an internal subset of the PSR areas that maintain the coldest temperatures and the lowest sublimation rates, considered most conducive to maintaining and accumulating water ice (Andreas 2006). A meter in diameter is postulated to be the lower diameter limit for cold trapped water ice, as constrained by low regolith thermal conductivity. Conductive heating from cold traps surrounding illuminated surfaces would otherwise heat and raise the internal temperature and sublimation rates for smaller area cold traps. Equatorward of 80° S, the PSRs generally become smaller, warmer, and less conducive to accumulating water ice (Mazarico et al. 2011; Hayne et al. 2021).

Surfaces may accumulate hydrogen-bearing volatiles from several potential sources, including outgassing from the lunar interior, as a remnant of the past comet and meteor bombardment, or from the ongoing production of water-ice molecules ($H_2O$) or hydroxyl ions ($OH^-$) that may be created by solar-wind proton interactions with lunar regolith silicates, which have important implications for hydrogen volatiles assessment and recovery efforts (Arnold 1979; Starukhina & Shkuratov 2000; Starukhina 2001, 2006; Saal et al. 2008; Crotts & Hummels 2009; Ong et al. 2010; Prem et al. 2015; Milliken & Li 2017).

Hydrogen-bearing volatiles may become broadly distributed during their poleward diffusion, which would yield a fraction of the migrating population being sequestered within PSR cold traps. Poleward migration of such volatiles may occur from micrometeorite impacts that eject volatiles with random trajectories. Diurnal surface thermal variation may drive volatiles away from high daytime temperatures, toward the terminator, the poles, and the PSRs (Crider & Vondrak 2000; Clark 2009; Pieters et al. 2009; Sunshine et al. 2009; Moores 2016). Evidence for migrating water-ice as gas molecules has been observed as exospheric plumes observed in near-infrared spectra of the Spectral Profiler on board the SELenological and ENgineering Explorer (SELENE)/Kaguya spacecraft (Ohtake et al. 2024). The plumes are attributed to subsurface sublimation events, and are observed diurnally, occurring primarily over non-PSR surfaces, an indication that non-PSR have higher loss rates relative to PSR, consistent with (Watson et al. 1961).

Solar incidence angles are extremely low toward the poles and small degrees of topographic slope variation create significant variation in the expected shadowed areal distributions, especially in contrasting poleward-facing slopes (PFS) and equator-facing slopes (EFS). South polar solar incidence angles are maximized diurnally when the Sun is at its greatest elevation toward the northern horizon at local noon. PFS aspects face away from the maximum solar incidence angle which yields more and larger shadowed areas, and colder surfaces, yielding a higher than expected PSR areal density. Conversely, the EFS aspects are rotated toward the maximum solar incidence yielding fewer and comparatively smaller shadowed areas, which are warmer, yielding a lower than expected PSR areal density.

Craters' hot EFS also become a biased source of secondary heating by reflecting long-wavelength radiation into crater basins and toward their respective PFS. Heating is reduced as a function of distance from the source, which creates a surface thermal gradient within craters and PSR basins evaluated in this study (Paige et al. 2010b; Mazarico et al. 2011; Schörghofer & Aharonson 2014; Moores 2016; Hayne et al. 2021). We postulate from the shared morphologies and temperature distributions within craters, PSRs, and cold traps that the PSRs' hydrogen distributions are similar and spatially scale with the PSR areas.

LEND CSETN observations poleward of 65° S showed hydrogen concentrations are greater toward PFS relative to significantly lesser concentrations observed on their respective EFS (McClanahan et al. 2015). A similar observation was observed by the Stratospheric Observatory for Infrared Astronomy (SOFIA), which detected molecular water concentrations that are locally enhanced on PFS, ranging from 100 to 400 parts per million (ppm), with lesser water concentrations on the EFS at Clavius crater [58°.2 S, 345°.6 E]. Importantly, water ice is uniquely identified by SOFIA's 6.1 $\mu$m emission band, which may disambiguate CSETN's hydrogen observations (Reach et al. 2023). The joint results may indicate a much broader distribution of water ice in PSRs and on small PFS cold traps than previously thought.

Infrared and ultraviolet reflectance techniques are sensitive to water ice and hydrogen-bearing volatiles in the regolith's top few microns (Pieters et al. 2009; Gladstone et al. 2010; Farrell 2019; Honnibal 2020). The reflectance techniques are also sensitive to surface thermal variation and solar irradiation. In comparison, neutron remote sensing methods maintain distinct advantages in the remote sensing of hydrogen-bearing volatiles in lunar polar conditions. Neutron techniques (a) enable regolith volume measurements, being sensitive to hydrogen concentrations within the surface top meter, (b) are nearly insensitive to surface thermal variation, and (c) are unaffected by solar irradiation variation. These properties make the technique ideal for comparing the neutron-emission flux from the cold PSRs and their surrounding warmer non-PSR surfaces (Feldman et al. 1991; Mitrofanov et al. 2010b). However, the interpretation of neutron observations is sensitive to hydrogen burial or layering conditions (Mitrofanov et al. 2010b; Lawrence et al. 2011b).

### 1.2. Lunar Neutron Studies

Orbital neutron spectroscopy techniques have a long history in geochemistry and hydrogen-bearing volatile studies of the Moon, Mars, Mercury, and asteroids (Feldman et al. 1998, 1999; Boynton et al. 2004; Goldsten et al. 2007; Mitrofanov et al. 2010a; Prettyman et al. 2012). Neutrons are emitted from planetary and small bodies after GeV-energy Galactic cosmic rays (GCRs) impact nuclei in the regolith. The spalled neutrons scatter throughout the regolith, dissipating their energies in subsequent collisions with regolith nuclei. Neutron energies are attenuated as a function of the neutron-scattering cross sections of regolith nuclei encountered before they escape the surface. Most neutrons are absorbed within the regolith as their energies reach thermal equilibrium. A fraction of the neutrons, which may originate up to a meter in depth, may escape the surfaces of airless small bodies, becoming detectable from orbit.

Water effectively moderates neutron energies because a neutron's mass is equivalent to a hydrogen nucleus (proton). A neutron loses, on average, a more significant fraction of its initial energy per collision with a hydrogen atom than the most common regolith elements. Where hydrogen is present in the regolith, its concentration yields a correlated suppression of the epithermal neutron-emission flux, described below. The evaluation of epithermal neutron suppression is a differential measure of the observed neutron count rates at the hydrogen-enhanced location relative to that of the neutron count rate at a region that is considered anhydrous. Monte Carlo neutron transport codes are used to quantify the conversion of epithermal neutron-suppression to hydrogen concentration maps (Feldman et al. 1991, 1998; Forster & Godfrey 2006; Lawrence et al. 2006; McKinney 2006; Mitrofanov et al. 2008, 2010a; Allison et al. 2016; Sanin et al. 2017).

Hydrogen may be a component of several potential hydrogen-bearing molecules and ion species (e.g., $H_2$, $H_2O$, $OH^-$, $CH_4$, and $NH_3$). However, oxygen is the third most common element and is nearly twice as common as carbon and nitrogen. This constraint indicates that most hydrogen-bearing volatile molecules in the carbon-poor lunar environment are likely water. This assertion is substantiated by ground-truth observations of the ejected plume of the spent rocket motor impact of the Lunar Crater Observing and Sensing Satellite (LCROSS; Colaprete et al. 2010). More than an order of magnitude more $H_2O$ molecules were detected by LCROSS than any other of the observed hydrogen-bearing molecules. The assertion is supported by the SOFIA water observation on PFS, and the SELENE observation of plumes of exospheric water (Reach et. al. 2023; Ohtake et al. 2024). From this evidence we adopt the term water-equivalent hydrogen (WEH) to describe the abundance of hydrogen atoms residing within the surface top meter that can influence the epithermal neutron leakage flux relative to an anhydrous surface, Section 2.

The uncollimated Lunar Prospector Neutron Spectrometer (LPNS) made the first definitive detection of polar epithermal neutron suppression in a latitude-dependent profile, which shows nearly symmetric suppression of the lunar epithermal neutron-emission flux within 15° of latitude of both poles (Feldman et al. 1998, 1999). However, the LPNS spatial resolution, 45 km FWHM ($\sigma = 19.1$ km), as observed in the low altitude LPNS mission phase of 30 km, yielded a $\pm 3\sigma$, 115 km diameter footprint that precluded the direct quantification of PSR hydrogen concentrations due to their substantially smaller areas (Maurice et al. 2004). Several image restoration studies have since partially corrected the LPNS instrumental blurring in its WEH maps by constraining the reconstructions with PSRs shadow models and field-of-view (FOV) factors. Those studies concluded that the LPNS instrumental blurring induced epithermal neutron spatial gradients in its maps that originate from the PSRs (Elphic et al. 2007; Eke et al. 2009; Teodoro et al. 2010, 2014; Wilson et al. 2018).

The LPNS findings were followed up a decade later in the design of the LRO mission instrument suite, with LEND, and specifically CSETN, designated to perform high-spatial-resolution mapping of the polar hydrogen-bearing volatiles (Mitrofanov et al. 2010a). LEND's signature instrument, CSETN, is designed to passively detect epithermal collimated lunar (CL) neutrons at a high spatial resolution. LEND was designed with a complement of eight $^3$He detectors and a scintillator detector designed to detect three lunar neutron energy ranges: thermal ($E < 0.4$ eV), epithermal (0.4 eV $< E < 300$ keV), and fast (300 keV $< E$; Mitrofanov et al. 2008, 2010a). CSETN, evaluated in this study, employs four of the $^3$He detectors. Observations of the other four $^3$He detectors are uncollimated, measuring thermal and epithermal neutrons, not evaluated in this study. CSETN's collimator is made of $^{10}$B and polyethylene that absorbs neutrons and discriminates those that have origins that are outside the collimated FOV. Open-ended tubes contain the $^3$He detectors within the collimator, and the tube lengths define their detector apertures, which subtend an angular FOV of 5.12°, as measured from the instrument boresight vector. Mitrofanov et al. (2008, 2010a) modeled the FOV from LRO's science mission altitude of 50 km, yielding an FWHM = 10 km and a $\pm 3\sigma$ footprint = 25.8 km. Our independent Geometry ANd Tracking version 4 (GEANT4) neutron transport modeling at 50 km altitude yields a slightly larger footprint, FWHM = 11.8, and a $\pm 3\sigma$ footprint = 30.6 km diameter.

A consequence of CSETNs' design is that it detects a bimodal neutron energy distribution. CL neutrons are observed at high spatial resolution. The CL neutrons pass freely through the four collimator apertures to be detected by CSETN's $^3$He detectors. LROs' attitude control typically keeps CSETN's collimator tubes aligned with the nadir-looking direction. CSETN detects uncollimated lunar (UL) neutrons because the collimator is not fully opaque to neutrons originating outside the collimated FOV. UL neutrons are primarily detected as a population of low spatial resolution, fast, and high-energy epithermal neutrons. UL neutrons become a detectable epithermal neutron population by CSETN after their energies are reduced by scattering from the spacecraft and collimator body. A background source of GCR protons is also detectable as spallation neutrons and charged particles produced from interactions with the spacecraft and collimator assembly. GCRs do not carry a lunar-induced contribution and are corrected in LEND's ground calibration, see Section 2 (Mitrofanov et al. 2010a; Litvak et al. 2012a, 2016; Livengood et al. 2018).

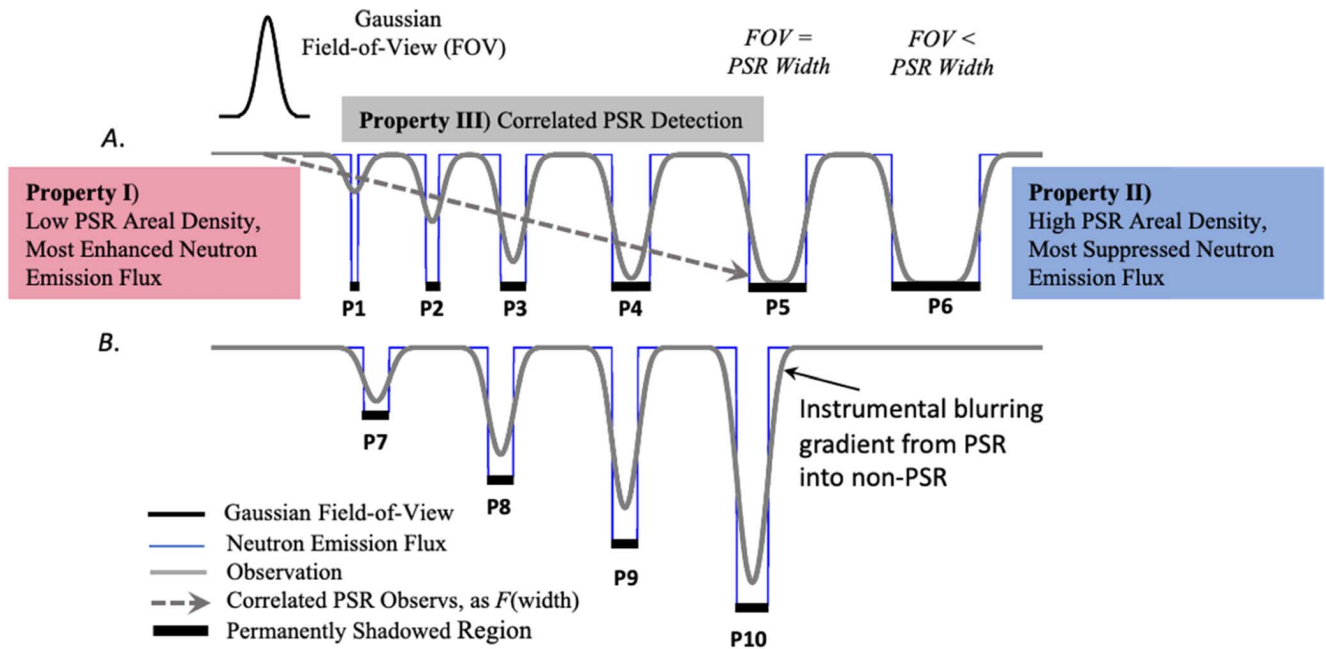

**Figure 1.** One-dimensional profiles of the "true" (blue) and observed (gray) neutron-emission flux for a series of PSRs, P1 to P6, detected by an orbiting neutron spectrometer. A fixed Gaussian FOV correlates (gray dashed arrow) the PSR's uniform neutron suppression as a function of their width. Profile (A) shows the correlated observation of uniformly suppressed PSRs caused by their increasing width. Profile (B) shows the partial detection of PSR widths when always less than the FOV. The model indicates nearly all PSRs will be fractionally detected as their areas are less than CSETN's footprint area.

Claims for CSETN's high spatial resolution have been the subject of vigorous and still unresolved debate. Several studies have presented widely contrasting evaluations of its performance. Early analysis of the LEND data showed significant hydrogen concentrations within several larger-area PSRs (Mitrofanov et al. 2010a, 2010b, 2012; Boynton et al. 2012; McClanahan et al. 2015; Sanin et al. 2017, 2019). However, several critical studies found that CSETN's detection of collimated neutrons was negligible (Lawrence et al. 2010, 2011a, 2022; Eke et al. 2012; Miller 2012; Teodoro et al. 2014). Before the present study, CSETN's detection of UL neutrons had been included in the LEND team's published PSR observations, thereby, several PSRs' observed hydrogen concentrations were overstated.

After nearly a decade of south pole observations, we reconsider these disparate evaluations and demonstrate evidence for both claims that CSETN is detecting UL neutrons and demonstrate its high-spatial-resolution detection capabilities through its correlated response to the PSR areal distribution. The correlation indicates the PSRs are similarly hydrogen enhanced and that their enhancement is a widespread phenomenon. To reach these objectives we quantify CSETNs' UL neutron suppression and subtract its mapped contribution from the CL neutron map using a spatial bandpass filter. The process yields CSETN's high-spatial-resolution CL maps. We validate the bandpass filter and methods by reviewing profiles of the PSRs neutron suppression before and after filtering in Section 3.1. The process is equivalent to background subtraction methods used to isolate photo peaks in spectroscopy applications (Evans et al. 2006).

*1.3. Hypothesis*

Our primary hypothesis is that the Moons' polar WEH distribution is biased toward the PSR's by their similar geomorphology and surface temperature distributions that yield similar WEH distributions, an effect that is independent of PSR spatial scale.. We postulate that random migration of water and/or hydroxyl volatiles distributes them uniformly to the surface. Their surface distribution becomes biased by the greater volatile loss rates from non-PSR relative to PSR surfaces (Watson et al. 1961; Arnold 1979). We assume that neutron-suppressed regions (NSRs) are synonymous with the PSRs. As this surface is observed by CSETN's collimated fixed-area FOV, the PSRs observed neutron suppression must be instrumentally blurred, and correlated by the proportional detection of their areas. The negative correlation is caused by the mixing ratio of neutron-suppressed PSR areas relative to relatively neutron-enhanced, non-PSR areas that are averaged within CSETN's footprint.

To illustrate this model, Figure 1, Profile (A), shows the "true" neutron-emission flux (blue), corresponding to a linear surface with uniform neutron-emission flux from non-PSR areas, embedded with uniformly suppressed neutron-emission flux from a series of increasing width PSRs, labeled P1 to P6 (thick black). The gray plot shows Profile (A)'s observation by a fixed-width Gaussian FOV. The negative correlation is observed in the series of proportional and increasingly suppressed PSR observations, P1 to P4 (gray arrow). P5 and P6 indicate the full detection of a PSR's "true" neutron suppression, where the FOV width ⩽ PSR width. Properties (I), (II), and (III) state the expected end member and transitional geophysical properties of the correlation.

Profile (B) shows that the neutron suppression is always a fraction <1.0 of the "true" suppression, P7 to P10, except where the PSR width ⩾ FOV width at Profile (A) P5 and P6. Instrumental blurring will also transfer the PSRs' neutron suppression as a suppression gradient that extends into adjacent non-PSR areas, P10.

Hypotheses to be tested.

$H_0$. The observed neutron-emission flux from the distribution of PSRs is not correlated with their a real density in CSETN's footprint. (Null Hypothesis.)

$H_A$. The observed neutron-emission flux from the distribution of PSRs is correlated with their a real density in CSETN's footprint.

To reject the null hypothesis the following Figure 1 properties must be jointly observed within CSETN's WEH, topography, and illumination maps.

*Property (I)*. The lowest observed hydrogen concentrations, i.e., CSETNs' greatest collimated counting rates, will occur at surfaces with the lowest PSR areal density (pink block on the left).

*Property (II)*. The greatest observed hydrogen concentrations, i.e., CSETN's lowest collimated counting rates, will occur at surfaces with the highest PSR areal density (blue block on the right).

*Property (III)*. The transition between the Property (I) and (II) areas is defined by instrumental blurring and CSETN's correlated observation of similarly WEH-enhanced (neutron-suppressed) PSR areas (gray block at the center).

Four additional predictions frame our Section 3.0 evaluations: (1) The lowest PSR areal density and highest collimated counting rates will generally be observed with increased distance from the PSRs. (2) WEH observations of the smallest-area PSRs will be negligibly correlated to their areas due to their inherently low signal-to-noise ratio. (3) The highest spatial gradients in the neutron-emission flux will be observed where PSR and non-PSR areas are adjacent. (4) PSRs' observed neutron suppression will be most degraded (reduced) where they are adjacent to areas of neutron-enhanced flux, Property (I) adjacent to Property (II). Evidence for predictions (3) and (4) is discussed in Section 4.

## 2. Methods

Section 2 reviews the analytical methods that were developed for this paper. Section 2.1 reviews the methods used to define the counting rate, statistics, and hydrogen map, as well as the spatial bandpass filter we use to isolate CSETN's high-spatial-resolution CL map from its UL neutron map.

More detailed history and background from the LEND team's prior studies, as well as its peer-reviewed calibration methods, can be reviewed (Mitrofanov et al. 2008, 2010a; Boynton et al. 2012; Litvak et al. 2012a, 2012b, 2016; Sanin et al. 2012, 2017; Livengood et al. 2018). This study correlates CSETNs' collimated WEH map to coregistered maps, including the Diviner radiometers' maximum temperature map, the Lunar Orbiter Laser Altimeter (LOLA) digital elevation map, a slope azimuth angle map, an averaged illumination map, and a binary PSR map, which is derived from the LOLA illumination map (Paige et al. 2010a; Smith et al. 2010; Mazarico et al. 2011). Figure A1 in the Appendix provides images of these maps.

### 2.1. Mapping

CSETN's count rate map CSETN_MK and statistical variance map CSETN_SK are derived by mapping the fully calibrated CSETN observations (Knuth 1998). To aid our following comparisons, we use the detector component rates and the neutron suppression to WEH conversions of Sanin et al. (2017). Our adoption of the Sanin et al. (2017) component rates and conversions for the present study, described below, is based on our comparable GEANT4 modeling of CSETN's GCR, CL, and UL count rate components (Allison et al. 2016; Su et al. 2018), see the Appendix. Sanin et al. (2017) break down CSETN's 5.1 counts s$^{-1}$ total count rate based on its first year of orbital operations, as 53.7% is attributed to GCR, 19.7% to CL neutrons, and 26.6% to UL neutrons, yielding component count rates of [2.74, 1.0, 1.36] counts s$^{-1}$, respectively.

We map CSETN's 1 Hz observations as a series of single detector observations. This method accounts for the loss of two detectors in 2011 May. The mapping differs from prior LEND team studies that had normalized combinations of valid detector observations, where $1 \leqslant n \leqslant 4$, emulating $n = 4$ detectors. The prior studies used the detectors' averaged count rates, established early in the mission, as proxies for observations that were deemed invalid (Boynton et al. 2012; Litvak et al. 2012a). However, since 2011 May, only two of CSETN's detectors have been in operation. Given the prior mapping method, all subsequent observations would require at least two of the four detectors to be set at constant values to emulate a four-detector observation. Such proxies would otherwise significantly distort the map counting rates and counting statistics in their accumulated maps. Based on this concern, we map only validated single detector observations. Given the GCR, CL, and UL component percentages, the total count rate for a single detector is 5.1 counts s$^{-1}$/4 detectors = 1.275 total counts s$^{-1}$ = [0.685, 0.250, 0.339] counts s$^{-1}$ per detector, respectively. GCR background is corrected in LEND's ground calibration.

CSETN cannot directly measure GCR and CSETN's UL and CL neutron counting rates. We model these rates and their negligible uncertainties with the GEANT4 neutron transport modeling software. Model inputs are GCR charged-particle spectra derived from Usoskin et al. (2011). The GCR charged-particle flux interacts with the physical and composition models of the spacecraft, LEND, and the lunar regolith to produce both the spacecraft-induced GCR background and the lunar neutron counting rate. Our model of the GCR charged-particle energy distributions is comprised of $10^6$ protons and alpha particles, which yielded $3 \times 10^7$ lunar neutrons. The uncertainty of the lunar neutron-emission flux is on the order of $10^{-4}$ neutrons cm$^{-2}$ s$^{-1}$ sr$^{-1}$. The uncertainty of the spacecraft-induced GCR background is on the order of $10^{-4}$. The lunar neutron-emission energy distribution, comprised of $5 \times 10^9$ neutrons, is projected from the lunar surface to interact with the LEND and CSETN models. The model tallies CSETN's UL and CL neutron captures, indicating 6000 and 9000, respectively. The UL and CL uncertainties are on the order of $10^{-5}$ counts s$^{-1}$. See the Appendix for more GEANT4 modeling discussion.

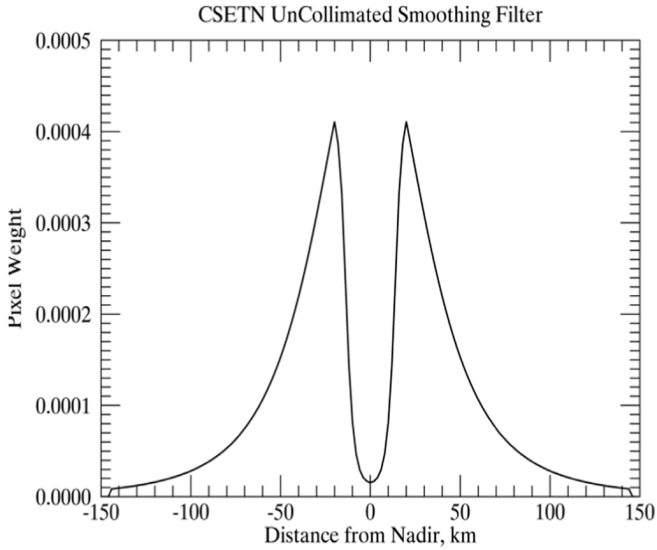

**Figure 2.** Shows a center profile of CSETN's UL, smoothing kernel, $W_{UL}$, centered over the nadir point at 50 km altitude. The kernel profile shows the relative pixel weights. The profile is symmetric around the nadir point = 0 km. The kernel is 145 × 145 pixels; each pixel is 2 km × 2 km. The cutoff of the UL kernel was set at ±4σ from nadir. The UL kernel was derived from GEANT4 modeling of collimated and uncollimated neutron contributions to CSETN.

Equation (1) derives the lunar neutron-suppression map, which is a function of the CL count rate after subtracting the sum of the GCR and UL rates. Equation (2) defines the total collimated neutron-suppression map $\epsilon$. The term is the Equation (1) map normalized to the anhydrous region counting rate, $\mu_{LB}$. The anhydrous region is defined before any analysis, and $\mu_{LB}$ represents the averaged CL neutron counting rate observed in the 65° S to 70° S latitude band. The choice of the anhydrous background region is important because $\mu_{LB}$ defines the frame of reference where wt% WEH = 0.0 (Lawrence et al. 2006; Sanin et al. 2017). Based on this setting, the results presented in this study do not include the Moon's estimated global average of 47 ppm (Sanin et al. 2017; Lawrence et al. 2022). Given the background counting rate, relatively positive neutron-suppression or "neutron-enhanced" areas impossibly suggest "negative wt% WEH" areas. We show that the neutron-enhanced regions are consistent with, and used in a proof of Property (I) of our model, discussed further in Section 3.1. The statistical uncertainty for the averaged count rate for the anhydrous background region $\mu_{LB}$ is negligible with a standard error of the mean count rate of $0.145/(1.38 \times 10^7)^{0.5} = 3.9 \times 10^{-5}$

$$L = \text{CSETN\_MK} - 1.024, \quad 1.024 \text{ counts s}^{-1} = (\text{UL} + \text{GCR rates}), \quad (1)$$

$$\epsilon = L/\mu_{LB}, \text{ Neutron suppression, counts s}^{-1}. \quad (2)$$

A spatial bandpass filter isolates CSETN's high-spatial-resolution CL neutron-suppression map. Spatial bandpass filters have a long history in electronics and digital image processing research. The technique isolates contrasting areas within images by quantifying and subtracting a local background image (Sonka et al. 1998; Gonzalez & Woods 2018). Such methods are variants of difference of Gaussian filters, which work by low-pass filtering (smooth) an input image $X$ using one of two, two-dimensional Gaussian smoothing kernels, $A$ and $B$, where Gaussian $B$ is of greater spatial width than $A$. The resulting image $X_A$ is the least smoothed and contains a superset of the spatial frequencies, including those of image $X_B$. Subtraction yields Image $X_C$, which contains an intermediate band (bandpass) of spatial frequencies, where image $X_C = X_A - X_B$.

We use an identical approach to isolate CSETNs' high-spatial-resolution CL neutron-suppression map. The Figure 2 UL kernel weights, $W_{UL}$, reflect the relative counting rates of UL neutron sources from pixels as a function of their distance from the nadir-looking direction at 0 km. Near nadir, the UL weights are small but not nonzero. The collimator faceplate also creates UL neutrons; however, the nadir-looking UL area is relatively small compared to the UL footprint. The UL weights are highest toward ±20 km, which reflects the likelihood of UL detection by scattering from the high volume of of collimator materials that impede those neutron detections. Further from nadir, the weights are reduced as the likelihood of such pixels being detected is reduced.

Equation (3), $\epsilon_C$ is the high-pass map, a smoothed neutron-suppression map containing both collimated and uncollimated contributions. The smoothing is performed using a two-dimensional Gaussian, $G_{CL}$, kernel with width = 11 km FWHM. The constant smoothing level is used to demonstrate widespread evidence of Property (III) and the transfer of enhanced PSR WEH as a spatial gradient toward lesser WEH distributed into non-PSR areas, in Section 3.5

$$\epsilon_C = \epsilon * G_{CL}. \quad (3)$$

Equation (4) defines the UL neutron-suppression map, $\epsilon_{UL}$, which is the low-pass component of the filter. The low-pass map bisects the neutron-suppression map $\epsilon$ and reflects a filter-weighted average of the neutron-emission flux. $\epsilon_{UL}$ defines the map of the UL suppression; however, its intensity is too high because the suppression of the $\epsilon_{UL}$ and the $\epsilon_{CL}$ maps (Equation (6)) should be complementary, as discussed below

$$\epsilon_{UL} = \epsilon * W_{UL}, \quad (4)$$

$$\epsilon_{UL} = ((\epsilon_{UL} - 1.0) * 0.5) + 1.0). \quad (5)$$

Equation (5) reduces the suppression intensity of the $\epsilon_{UL}$ map for its subtraction from the Equation (6) map. The scale term = 0.50 was set from the observation that both LPNS and LENDs SETN instruments observe an ∼5% polar neutron suppression (Feldman et al. 1998; Mitrofanov et al. 2010b). The scale term setting is cross validated in the Cabeus-1 profile study showing the complementary CL and UL components, Section 3.1.1.

Equation (6) isolates the high-spatial-resolution collimated neutron-suppression map $\epsilon_{CL}$ by subtracting the scaled, uncollimated neutron-suppression map, $\epsilon_{UL}$. The collimated neutron-suppression at the pole after the bandpass is ∼5%. The map includes the collimated neutron suppression of both individual PSRs and background neutron suppression, postulated from Figure 1. The background neutron suppression should correlate to the increasing poleward density of small, similarly hydrogen-enhanced cold-trap areas (less than 2 km wide pixels; Hayne et al. 2021). The bandpass filter and the derived scale term are discussed further in Section 3.1

$$\epsilon_{CL} = (\epsilon_C - \epsilon_{UL}) + 1. \quad (6)$$

Equation (7) transforms the collimated neutron-suppression map $\epsilon_{CL}$ to its corresponding wt% WEH map, $C_{WEH}$, where parameters are $a, b, c = [1.2, 0.06, -0.51]$ (Sanin et al. 2017).

The method yields a continuous but nonlinear transformation of the collimated neutron-suppression map to wt% WEH

$$C_{WEH} = (-a + (a^2 + 4b*(\epsilon_{CL}^{1/c} - 1))^{0.5})/2b. \quad (7)$$

Equation (8) derives the standard error $\delta$ map, where each pixel is a Poisson-distributed random variable. The count rate variance map, CSETN_SK, is normalized by the observation counts map by the number of observations in each pixel, $n$

$$\delta = \sqrt{CSETN\_SK/n}. \quad (8)$$

Equation (9) adds the standard error in quadrature to reflect the subtraction of the UL random variable (map), Equation (6). Note that Equation (9) assumes that the UL and CL count rate components are independent random variables. However, they are not because their expected counting rates are positively correlated to the GCR variation during the LRO mission. Their positive correlation implies that the statistical uncertainty $\sigma$ map is an upper-bounds statement.

$$\sigma = \sqrt{\delta^2 + \delta^2}. \quad (9)$$

Equation (10) defines the derivation of Spearman's correlation coefficient used to correlate the four Figures 5(a)–(d) profiles. The studies compare their respective (2) collimated neutron suppression to (3) maximum temperatures. Spearman's correlation coefficient method assumes the two observation variables are monotonic and operates on the relative rank of their observations. The two distributions of rankings are paired for analysis (Press et al. 1992)

$$\rho_{sm} = \frac{\sum_{i=1}^{n}(R_i - \mu_R)(S_i - \mu_S)}{\sqrt{\sum_{i=1}^{n}(R_i - \mu_R)^2}\sqrt{\sum_{i=1}^{n}(S_i - \mu_S)^2}}. \quad (10)$$

Here $\rho_{sm}$ is the Spearman's correlation coefficient, $R_i$ is a sample from rank variable $R$, $S_i$ is a sample from rank variable $S$, $\mu_R$ and $\mu_S$ each define the average of the rankings in each variable $R$ and $S$, respectively, and $n$ is the number of samples in each variable.

The significance $t$ of the $\rho_{sm}$ correlation is tested by relating $\rho_{sm}$ to the Student's $t$-distribution with $n-2$ degrees of freedom, Equation (11)

$$t = \rho_{sm}\sqrt{\frac{n-2}{1-\rho_{sm}^2}}. \quad (11)$$

Equation (12) defines the coefficient of determination $R^2$, which measures the goodness of fit for the correlated neutron and maximum temperature observations in Section 3.2. $R^2$ defines the proportion of variation in the observation predicted by the statistical model. $R^2 = 1$ indicates the statistical model entirely explains the observation variance

$$R_2 = 1 - \frac{\sum(y_i - \hat{y}_i)}{\sum(y_i - \bar{y})}. \quad (12)$$

Here $y_i$ is a dependent observation in each random variable PSR, where $\hat{y}_i$ is the prediction from the linear model, and $\bar{y}$ is the mean of the set of PSR observations.

Image region growing is a software method used extensively in this study to segment, isolate, and evaluate independent map areas (Sonka et al. 1998). It is an image processing technique that uses predefined criteria for initializing and stopping the aggregation of contiguous image pixels into independent areas, $A_i$. All pixels in each area are given the same unique numerical identifier. The segmentation enables the evaluation of all pixels within $A_i$.

We use image growing software to segment a binary PSR map into individual PSR areas, as shown in the Appendix, Figure A1. We assume PSR areas are of circular shape and Equation (13) converts their areas to their diameters, $d_i$

$$d_i = (A_i/\pi)^{0.5} * 2. \quad (13)$$

### 3. Results

Section 3 reviews results derived from CSETN's collimated mapping, poleward of 82° S. Section 3.1 presents CSETNs' maps and demonstrates its neutron-enhanced areas are consistent with Property (I) of our Figure 1 model. Section 3.1.1 validates the map processing pipeline, the bandpass filter, and WEH derivations. Section 3.2 shows the correlation of the WEH within the Haworth, Shoemaker, and Faustini PSR basins to their internal maximum temperature distributions. Section 3.3 shows the latitude-dependent enhanced WEH contrast in PSRs relative to non-PSR surfaces. Section 3.4 shows the isolated PSR WEH distribution, as correlated to their diameters. Section 3.5 shows the postulated correlation of the PSRs' WEH response that validates the detection of Properties (I), (II), and (III) areas.

#### 3.1. CSETN Collimated WEH Map Reviews and Validation

Figures 3(a)–(d) show CSETN's south pole stereographic maps, as derived from 10.5 yr of observations spanning 2009 July 2 to 2019 December 15, ⩾82° S. Figure 3(a) shows CSETN's collimated WEH concentration map after applying the spatial bandpass filter, Equation (7). Property (II) from our Figure 1 model and CSETN's high-spatial-resolution capability is demonstrated in its observation of the four greatest WEH concentrations that are coincident with several of the south pole's largest and most detectable PSR areas within the Cabeus-1, Haworth, Shoemaker, and Faustini craters (yellow). The maximum south pole WEH concentration is observed at the elongated ∼14 km × ∼21 km Cabeus-1 PSR, which indicates a narrow narrow response that is strongly aligned with its PSR centroid and is spatially isolated from the other large-area PSRs (Figure 3(d)). Cabeus-1 is importantly surrounded by a low-illumination area and a cluster of smaller PSRs, which may broaden and increase its WEH response, as detailed in the following sections. CSETN's high-spatial-resolution capability is further evidenced in its detection of comparatively less WEH observed at the two 20 km wide, irregularly illuminated ridges that span the gaps between the Haworth and Shoemaker, as well as Shoemaker and Faustini PSRs. Note that the "1" in Cabeus-1 indicates the largest-area individual PSR within the Cabeus crater.

The statistical uncertainty $\sigma$ map is derived from neutron counting statistics, as measured in 2 km pixels (upper bounds), Equation (9), Figure 3(b). Figure 3(c) shows CSETN's scaled uncollimated WEH concentration map, as removed by the bandpass filter, Equations (3)–(6). Figure 3(d) shows the LOLA digital elevation map (gray) with four corresponding longitude profiles (yellow). The profiles are 150 km long and bisect the Cabeus-1, Haworth, Shoemaker, and Faustini PSRs through their most WEH-enhanced pixels. The profiles are used to validate our methods and to demonstrate the correlation of CSETN's collimated and uncollimated maps to coregistered LOLA topography and maximum temperature maps, Section 3.1.1. CSETN's full-width 30 km diameter collimated

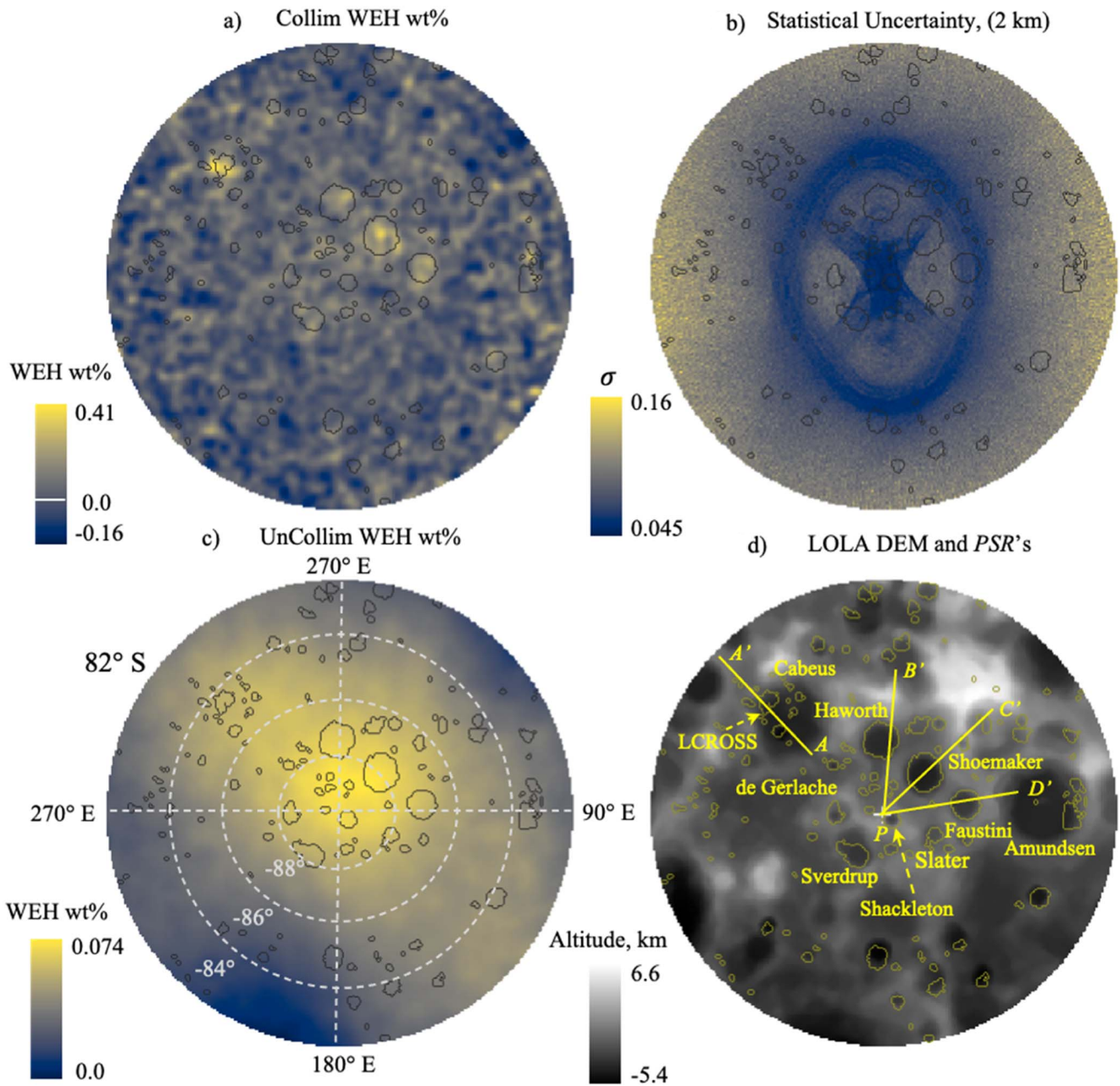

**Figure 3.** CSETN south pole maps after 10.5 yr of observations for latitudes poleward of 82° S. (a) CSETN's collimated wt% WEH map. (b) CSETN's upper-bounds statistical uncertainty map, in units of counts s$^{-1}$, Equation (9). The (dark blue) circle and the cross pattern indicate high coverage regions = low statistical uncertainty. (c) CSETN's uncollimated wt% WEH map, Equation (5). (d) Coregistered LOLA topography map (gray), with altitude in units of km deviation from the volumetric mean lunar radius = 1737.4 km. Outlines are given for PSRs with areas that exceed 20 km$^2$ (olive). The A′, B′, C′, and D′ longitude profiles bisect the most strongly WEH-enhanced locations within the Cabeus-1, Haworth, Shoemaker, and Faustini PSRs, respectively.

footprint is approximated by Faustini crater's 29 km diameter PSR (Figure 3(d)).

The Figure 3(a) map shows a large degree of WEH spatial variation, part of which is due to pixels with varying neutron count rates due to statistics. However, we demonstrate that a significant source of this variation is attributed to the spatial distribution and observation of similarly WEH-enhanced PSR areas. Neutron-enhanced areas (dark blue) impossibly suggest "negative WEH," i.e., below 0.0 wt% WEH. As noted in Section 2, the neutron-enhanced pixels simply indicate that their collimated counting rate exceeds that of the anhydrous region, Equation (2). The average of all neutron-enhanced areas poleward of 85° S is $\epsilon_{CL} = 1.021$, which indicates the possible need for a minor correction of the uncollimated background by reducing the Equation (5) scale term below 0.5, or the need to redefine the location of the a priori defined anhydrous region. Implementing either approach will yield slightly increased polar PSR-neutron-suppression and WEH concentrations, as demonstrated below.

For the course of Section 3, we demonstrate that the neutron-enhanced regions embody Property (I) of our model. Neutron-enhanced regions also constitute a new line of evidence that demonstrates CSETN's high-spatial-resolution observing capabilities. The study maps where there is a low expectation of water ice deposits and quantifies why.

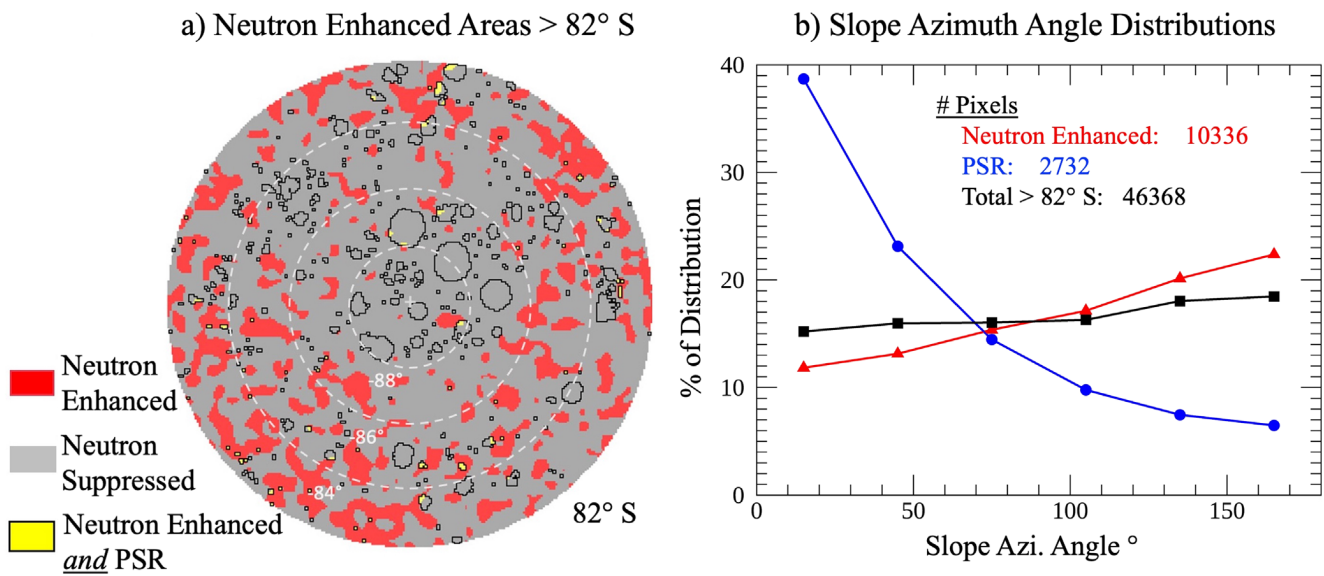

**Figure 4.** Characterization of neutron-enhanced areas, >82° S. (a) The map locates the neutron-enhanced areas (red), indicating there is an expectation of less WEH than the anhydrous region. Neutron-suppressed areas (gray) locate where there is an expectation of enhanced WEH relative to the anhydrous region. Yellow areas are permanently shadowed and neutron-enhanced pixels. (b) Profiles of slope azimuth angle distributions for the contrasting neutron-enhanced distribution from panel (a) (red triangles) relative to the PSR (blue circles) and all pixels poleward of 82° S (black squares). (a) PSRs are outlined in black.

Figure 4(a) shows the segmentation of neutron-enhanced (negative WEH, red) and neutron-suppressed areas (positive WEH, gray) from the Figure 3(a) map, classified relative to the 0.0 wt% WEH threshold. The 0% WEH threshold corresponds to the anhydrous region's neutron counting rate from Equation (2). Thereby, neutron-suppressed and neutron-enhanced areas constitute 78% ($1.44 \times 10^5$ km$^2$) and 22% ($4.13 \times 10^4$ km$^2$), respectively, of the total area $1.85 \times 10^5$ poleward of 82° S. CSETN classified 94.5% ($1.03 \times 10^4$ km$^2$) of the total PSR area ($1.09 \times 10^4$ km$^2$) as being neutron-suppressed area. The remaining 5.5% ($6.2 \times 10^2$ km$^2$) is PSR area observed within neutron-enhanced areas. The map shows 145 independent neutron-enhanced regions, most of which surround the larger PSRs which infrequently overlap them, indicating the regions have a low PSR areal density. The few PSR pixels that are also neutron-enhanced are (yellow).

Three tests quantify the regions low PSR areal density: (1) that its fraction of PSR-neutron-enhanced area is lower than expected; (2) that the expected area of individual PSRs within the neutron-enhanced areas are smaller than expected, i.e., low signal-to-noise ratio; and (3) that the expected locations where neutron-enhanced areas are observed are biased toward the craters' warm EFS, and away from PFS, i.e., where PSR areas are concentrated. Image region growing software is used to derive the neutron-suppressed PSRs, the neutron-enhanced, and the neutron-enhanced-PSRs areas, Equation (13) (Sonka et al. 1998).

Test (1) PSR areas constitute 1.5% of the map's neutron-enhanced area, which is less than the respective 7.6% observed within the map's total neutron-suppressed area (Mazarico et al. 2011). Thereby, neutron-enhanced areas have a relatively lower PSR areal density than the map's neutron-suppressed area.

Test (2) contrasts the two segmentations PSR diameter distributions. The mean PSR diameters and standard deviations within the neutron-enhanced and neutron-suppressed regions are $3.3 \pm 1.5$ km and $4.7 \pm 4.9$ km, respectively. The expected diameter range of neutron-suppressed PSR diameters are negligibly detectable PSRs with areas that are ~1% of CSETN's collimated footprint area of 707 km$^2$.

Statistical tests of the segmentation's two PSR diameter distributions indicate they are significantly different, based the nonparametric Kolmogorov–Smirnoff (K-S) test (Press et al. 1992). The K-S test reports the parameter $D$, which defines the K-S test statistic and states the maximum distance between the two normalized cumulative distribution functions. The test also reports the probability $p$, that the two distributions are drawn from the same parent population. Small $p$ below the defined significance threshold, $p < 0.01$, indicates the distributions are significantly different. Test (2) reported K-S test parameters as $[D, p] = [0.45, 2.2 \times 10^{-10}]$. Thereby the null hypothesis is rejected. Tests (1) and (2) show that (a) the neutron-enhanced areas have a relatively smaller PSR areal percentage than the neutron-suppressed areas, and (b) that the PSR-neutron-enhanced areas are aggregated into significantly smaller PSR diameters, relative to those same measures within the neutron-suppressed area.

To demonstrate the contrasting geophysical properties of the neutron-enhanced and PSR areas, we tested the regions' coregistered pixel distributions using the corresponding slope azimuth angle map (McClanahan et al. 2015). Slope azimuth angle is a measure of a pixel's topographic slope aspect orientation, as defined by the angular deviation of its slope normal vector from the south pole direction. The slope azimuth angle ranges between 0° at PFS, when the slope normal vector points to the south pole, and is 180°, when it is toward the equator. East- and west-facing slopes are coaveraged. Figure 4(b) contrasts the percentage profiles of the three, slope azimuth angle distributions, as binned into six groups of 30°: (a) all neutron-enhanced pixels in Figure 4(a) (red triangles), (b) all PSR pixels >82° S (blue circles), and (c) all pixels >82° S (black squares). The neutron-enhanced profile is biased toward the warm EFS, indicating nearly twice as many EFS pixels as its PFS. The all-observed distribution shows a relatively uniform distribution. The PSR distribution is as expected, strongly biased toward the PFS. The three slope azimuth angle distributions show the expected maximum temperature contrasts by their means and standard deviations: (a) $242 \pm 34$ K, (b) $93 \pm 34$ K, and (c) $231 \pm 49$ K, respectively.

The nonparametric K-S test evaluates if the slope azimuth angle distribution of the (a) neutron-enhanced distribution is significantly different from the (c) all-observed distribution (Figure 4(b); see also Press et al. 1992). The reported K-S test parameters are $[D, p] = [0.07, 1.6 \times 10^{-36}]$. The result indicates that the (a) neutron-enhanced distribution is significantly different than the (c) all-observed distribution. The K-S test indicates a significantly greater difference for the (a) neutron-enhanced versus the (b) PSR distributions, $[D, p] = [0.37, 0.0]$.

In summary, the Figures 4(a) and (b) results show that CSETN has isolated the contrasting Property (I) and (II) areas. Property (I) neutron-enhanced areas are relatively anhydrous, they have a low PSR areal density, they have smaller than expected, less detectable PSR areas, and they are significantly biased toward the warm EFS. Property (II) neutron-suppressed areas indicate enhanced hydrogen concentrations, where there exists a higher PSR areal density; these areas are strongly biased toward the PFS and colder than expected temperatures. Note that we examined the possibility of a sampling bias and performed the same K-S tests on distributions selected from several smaller latitude band ranges, poleward of 82° S. The tests all validated the larger study and its conclusions.

Section 4 reviews the largest neutron-enhanced area poleward of 88° S and how strongly contrasting illumination distributions and instrumental blurring have reduced several important PSRs' WEH observations, e.g., at Shackleton and Slater craters.

### 3.1.1. Permanently Shadowed Region Profiles

The Figure 3(d) PSR-neutron-emission flux profiles of Cabeus-1, Haworth, Shoemaker, and Faustini are correlated with their coregistered topography and maximum temperature profiles in Figures 5(a)–(d). The study isolates CSETN's high-spatial-resolution detection of both Property (I) and (II) areas with their geophysical context. The isolation of the CL map by the spatial bandpass filter is demonstrated and our processing pipeline is validated.

For each PSR, the coregistered profiles are presented in row order 1 to 4. Row (1) shows CSETN's total lunar neutron suppression, after mapping and shows its uncollimated neutron suppression (red), as $\epsilon_{UL} + \epsilon_{CL}$ (Equation (2)). No neutron suppression = 1.0 (black dashed line) implies no difference in neutron suppression relative to the anhydrous background. Row (2) shows collimated neutron suppression $\epsilon_{CL}$ after the bandpass filter and subtraction of the $\epsilon_{UL}$ suppression, Equation (6). The $C_{WEH}(\epsilon_{CL})$ converts collimated neutron suppression to wt% WEH, Equation (7). Row (3) gives coregistered Diviner maximum temperature profiles. Gray shows the 2 km pixel resolution profiles. To correlate the differing instrument spatial resolutions, the gray dashed plot shows the smoothed maximum temperature profile after convolution with CSETN's collimated smoothing kernel $G_{CL}$, Equation (2). Row (4) shows coregistered LOLA elevation profiles, which demonstrate the correlation of neutron suppression to topography, showing PSRs (thick black) and non-PSRs areas, as well as contrasting PFS to EFS.

No difference is observed for the four PSR studies in comparison to their collimated neutron suppression $\epsilon_{CL}$, before or after applying the bandpass filter, Row (1) to Row (2).

Quantifying the individual PSR WEH concentrations will require additional analyses not covered in this study. The process will require isolating PSR-specific areas and their WEH contributions from their local background WEH concentrations. The Spearman correlation coefficient $\rho_{sm}$ quantifies the significance of the relationship between the (3) smoothed maximum temperature profile and the (2) collimated neutron-suppression profile, reported in Table 1 at the end of Section 3.1.1. Statistical methods were reviewed at the end of Section 2.

Table 1 lists the Figure 5 profiles' measures of their maximum neutron-suppression latitudes and longitudes, the observed WEH at those locations, maximum temperatures, and testing results.

The joint probability that the south poles' four most neutron-suppressed locations are coincident with high PSR areal densities is defined by our 2 km pixel mapping, where the PSR area constitutes 5.9% of the total area, poleward of 82° S. The probability that just one of these PSRs could have been randomly identified by CSETN as a neutron-suppressed area is 0.059. Thereby, the joint probability that all four strongly neutron-suppressed areas might randomly coincide with PSRs = $(0.059)^4 = 1.2 \times 10^{-5} < p = 0.01$, where $p$ is the rejection threshold. Since a substantial fraction of the PSRs' areal distribution is too small for CSETN to detect, the PSR areal fraction of the map is reduced, and the joint probability these larger-area PSRs were randomly detected is demonstrably reduced. Therefore, the null hypothesis that these joint detections could have all randomly occurred from neutron counting statistics is rejected.

Figures 5(a1)–(a4) show the 150 km long longitude profile through the Cabeus-1 PSR, along 311° E, first directly observed over a decade ago in Mitrofanov et al. (2010b) and Sanin et al. (2012). Cabeus-1 shows an anomalously enhanced maximum south pole neutron suppression, $\epsilon_{CL} = 0.82$, located adjacent to and on the northwest edge of the PSR (Table 1 and Figure 5(a2)). The maximum suppression location is elevated in altitude, over 2 km above the crater basin and on Cabeus craters' PFS. That location is $\sim$11 km from Cabeus-1 PSRs' coldest location in its basin, Figures 5(a3)–(a4). The 40 km wide neutron suppression is centered over the 13 km wide Cabeus-1 PSR. This result constitutes a strong line of evidence that CSETN's collimator is functioning as designed with a positive counting rate (Figures 5(a2)–(a4)).

The Equation (5) scaling term is cross validated by showing that the collimated and uncollimated neutron suppression are complementary components. This result is demonstrated by the near alignment of the uncollimated suppression profile and the most neutron-enhanced segments of the collimated suppression. The strong alignment of the WEH response to the Cabeus-1 PSR area is shown in comparing Figures 5(a2) to (a4). Figures 3(d), 4(a) show the Cabeus-1 map with its surrounding cluster of smaller area PSRs, whose WEH response may be complementary and spatially insulating by providing a >15 km buffer from the contrasting flux of several neutron-enhanced areas, Section 4 discussion.

Cabeus-1's most neutron-suppressed location and anomalous suppression magnitude suggest that a second process is complementing the region's observed WEH concentration. The maximum suppression location is $\sim$10 km from the ground-truth water-ice observation of 5.6% $\pm$ 2.9% by mass observed with a maximum excavated mass of 155 kg in the ejected plume of the LCROSS' spent Centaur rocket motor impact [84°.68 S, 311°.31 E] (Colaprete et al. 2010), Figures 3(d), and 5(a4). The infrared and ultraviolet instruments on board the LCROSS shepherding satellite observed the impact. Cabeus-1 was detected in early LEND CSETN observations that had included uncollimated neutrons (Mitrofanov et al. 2010a). LRO's Lyman-Alpha Mapping Project (LAMP) detected 140 kg of molecular hydrogen in the ejected plume (Gladstone et al. 2010). The plume was also observed in Earth ground-based observations to have 6.3% ± 1.3% water by mass (Stryker et al. 2013).

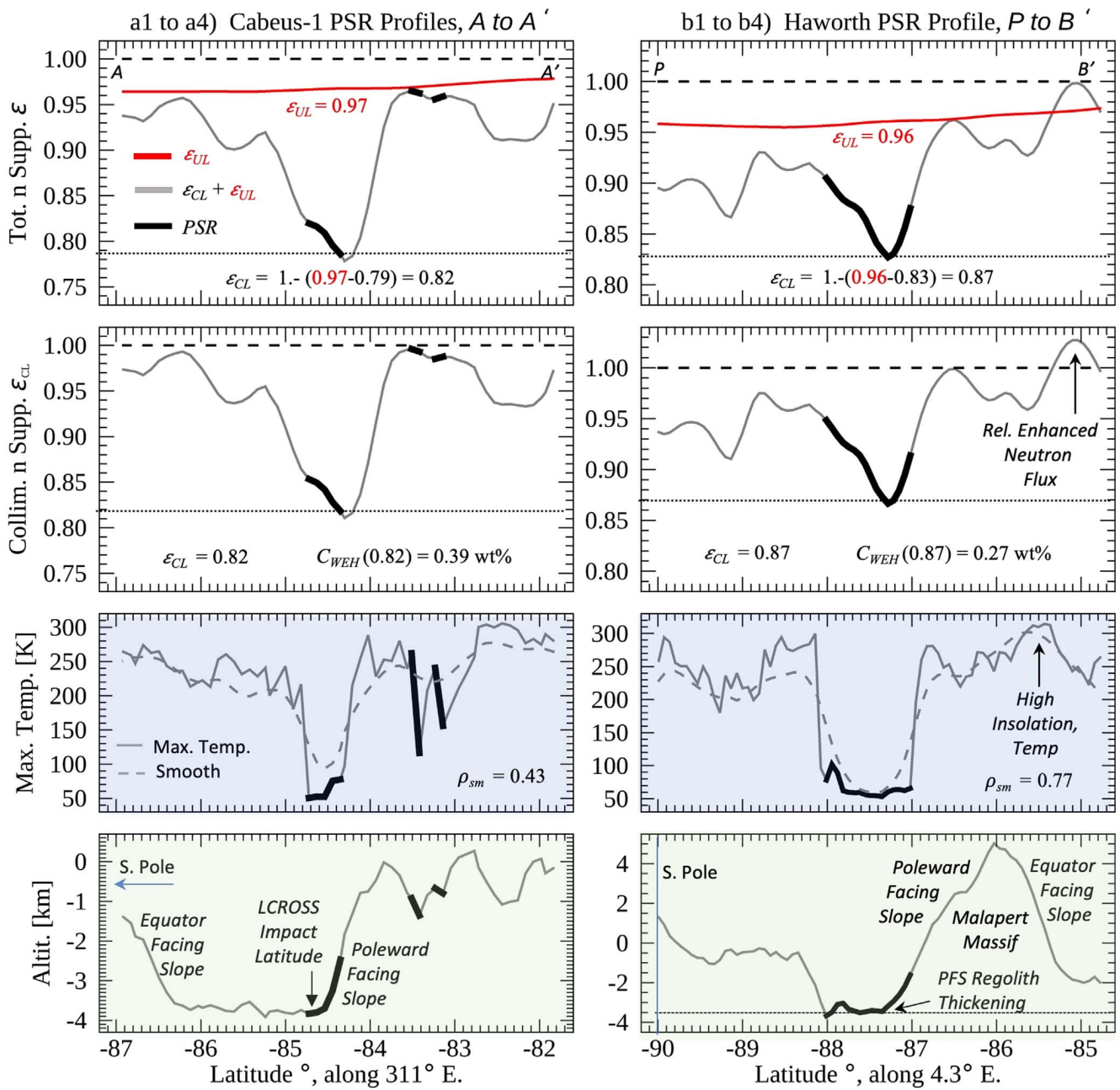

**Figure 5.** Comparison of the CSETN PSR profiles before and after the bandpass filter, plus the derivation of collimated wt% WEH. Columns show the PSR profiles of (a1)–(a4) Cabeus-1, (b1)–(b4) Haworth, (c1)–(c4) Shoemaker, and (d1)–(d4) Faustini from Figure 3(d). Plots within columns show coregistered PSR profiles in row order (gray): (1) total neutron suppression, after mapping, where uncollimated neutron suppression is in red. Black dashed lines show the anhydrous level, where no suppression = 1.0; (2) collimated neutron suppression after bandpass filter, with wt% WEH conversion; (3) Diviner maximum temperature profiles and smoothed maximum temperature (gray dashed), where the background is shown as a blue line; and (4) LOLA altitude in units of km deviation from the 1737.4 km mean lunar radius (green background). Coregistered PSR pixels are in thick black. Dotted black lines show the altitude baseline and evidence for regolith thickening toward the PFS.

Cabeus-1's maximum neutron-suppression location near the PSRs' northwestern edge closely coincides with the image-reconstructed LPNS maximum WEH location (Elphic et al. 2007; Figure 2). Regional epithermal neutron-emission flux gradients from the LPNS established evidence for past polar wander and a southern paleo-pole location at [84°.1 S, 309°.4 E] (Siegler et al. 2016). Due to the LPNS broad FOV, its maximum WEH location is negligibly different from CSETN's, just 11.5 km to the northwest. The profile's most neutron-enhanced locations north and south of Cabeus-1's PSR have warmer 250 K surfaces that are only weakly neutron suppressed, $\epsilon_{CL} < 1.0$, Equation (2), Figures 5(a2)–(a4). Sections 3 and 4 discuss Cabeus-1's anomalous maximum neutron suppression location.

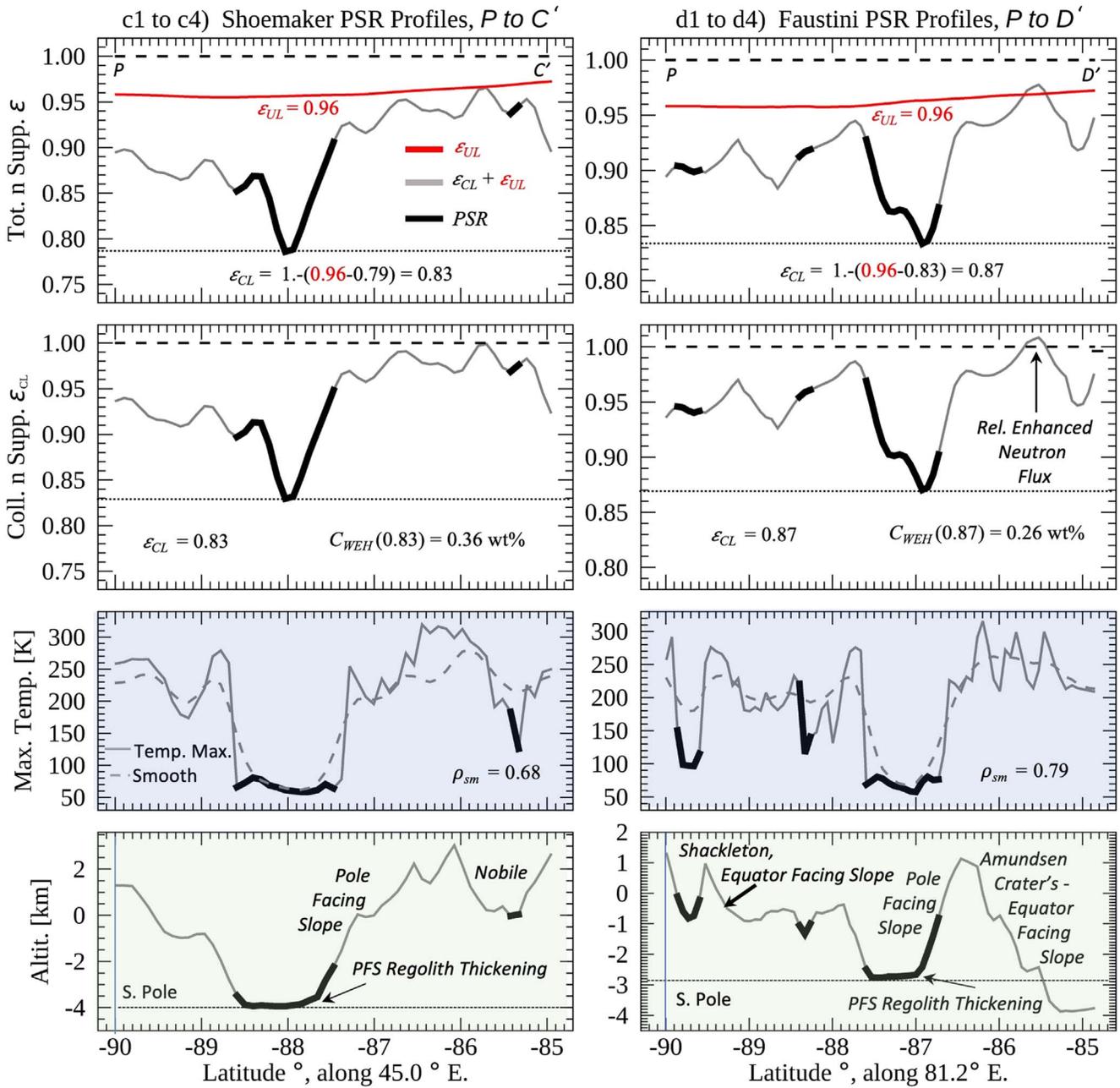

Figure 5. (Continued.)

Table 1
Figures 5(a)–(d), Permanently Shadowed Region Maximum Water-equivalent Hydrogen, Locations, and Correlated Profiles

| PSRs | Max. Collim. Supp. Latitude | Max. Collim. Supp. Longitude | Collim. Supp. | Collim. WEH wt% | Max. Temp. (K) | Spearman $\rho_{sm}$ | Signif. $t$ |
|---|---|---|---|---|---|---|---|
| Cabeus-1, $A$ to $A'$ | −84.35 | 311.78 | 0.82 | 0.39 | 78 | 0.43 | 6.6E−04 |
| Haworth, $P$ to $B'$ | −87.28 | 4.31 | 0.87 | 0.27 | 60 | 0.7 | 3.9E−13 |
| Shoemaker, $P$ to $C'$ | −87.98 | 43.67 | 0.83 | 0.37 | 59 | 0.68 | 9.6E−09 |
| Faustini, $P$ to $D'$ | −86.92 | 81.24 | 0.87 | 0.26 | 73 | 0.79 | 4.8E−18 |

**Note.** Column (1): PSR name and profile ID. Columns (2) and (3): latitude and longitude of the PSR's maximum collimated suppression. Column (4): PSR's maximum collimated suppression. Column (5): PSR's maximum collimated WEH in units of wt%. Column (6): maximum temperature at the maximum suppression location. Columns (7) and (8): profile Spearman correlation coefficient $\rho_{sm}$ and statistical significance $t$ for collimated neutron-suppression versus coregistered smoothed maximum temperature profile, respectively, Figures 5(a3)–(d3).

Figures 5(b1)–(b4) show the 150 km long profiles through the 35 km diameter Haworth PSR, through its maximum neutron-suppression location, $\epsilon_{CL} = 0.87$, along 4°.3 E longitude. The profiles demonstrate evidence corresponding to both Property (I) and Property (II) areas (Figure 1). The Property (I) area of neutron-enhanced flux occurs north of the Malapert Massif. (Figure 4) Property (II) is defined by the neutron-suppression within the Haworth PSR. The PSR shows evidence for heterogeneous WEH distributions within its area by its asymmetric neutron-suppression gradients and maximum suppression location north of its PSR's centroid. This location corresponds to the base of Haworth's PFS coinciding with a maximum surface temperature = 60.3 K.

Haworth's neutron-enhanced location occurs at the EFS base of Malapert Massif. From its EFS base, the slope rises a total of 7.5 km, with an average slope of 17°.4. Figure 5(b3) shows the ridge's high elevation, high slope degree, and EFS orientation, which cause high insolation, with 320 K maximum temperatures toward the ridges upper elevations, with lesser 220 K temperatures near its EFS base. An examination of the Malapert EFS with the LRO Camera (LROC) Lunar Quickmap system indicates (b) there are few PSRs with ⩾1 km diameter for areas 30 km north of the ridge which encompasses several tens of kilometers east and west of the base, with conditions consistent with the Property (I) low PSR areal density. The EFS base and any constituent small cold traps may also be warmed by secondary heating from long-wavelength radiation, reflected at up to a 17°.4 elevation angle from the hot upper EFS to the slope base, creating a WEH depleted surface.

It is important to note that the more neutron-enhanced location is aligned with the relatively cooler EFS base, and not Malapert Massif's hot upper elevation EFS, Figures 5(b2)–(b4). Theoretical studies have found that epithermal neutrons are weakly sensitive to surface temperature variation (Sanin et al. 2017). The alignment with the cooler slope base reduces the possibility of a temperature-dependent explanation for the neutron flux distribution. Note the relative thickening at the base of Haworth's PFS, which may be consistent with enhanced water-ice concentrations in the regolith subsurface, Figure 5(b4); see also Rubanenko et al. (2019).

Figures 5(c1)–(c4) shows the 150 km long Shoemaker PSR longitude profiles, along 45° E, adjacent to its maximum neutron suppression, $\epsilon_{CL} = 0.83$. The suppression location occurs at the base of its broadened PFS (Figure 5(c4)). Shoemaker's neutron suppression was claimed to have significant neutron suppression in Sanin et al. (2012) and Boynton et al. (2012). Based on the spatially sharp 30 km wide neutron, such a claim could only be made if CSETN's collimator provides high spatial resolution, indicating a positive CL neutron count rate. The study showed a bimodal detection, with a sharper CL neutron suppression, complemented by a broader regional neutron suppression, consistent with the expected spatial width of the UL neutron suppression. In considering Figure 1, cases P1 to P5, the spatially sharp neutron suppression is an indication that CSETN only fractionally detects the "true" neutron suppression at Shoemaker, which indicates a lower-bounds WEH detection. The maximum neutron-suppression location occurs at the cold 59.5 K base of Shoemaker's PFS. Note the relative thickening at the base of Shoemaker's PFS, which may be consistent with enhanced water-ice concentrations in the regolith subsurface (Figure 5(c4); again see Rubanenko et al. 2019).

Teodoro et al. (2014) revisited Shoemaker's 45° E longitude profile from the Boynton et al. (2012) study and evaluated the first 19 months with self-calibrated observations. That study concluded that the detection was likely due to statistical noise. The study further claimed that CSETN's collimated count rate is negligible. The present study reviews the Shoemaker profile after the number of PSR observations nearly doubled in the ensuing decade. Yet, Shoemaker's PSR maintains the second-greatest neutron suppression in the southern hemisphere, so its continued strong suppression is likely not a statistical aberration, which our Section 3 results further substantiate.

Figures 5(d1)–(d4) show the 150 km long Faustini longitude profile along 81°.2 E, with a maximum neutron suppression, $\epsilon_{CL} = 0.82$. As was shown for Haworth, the profile shows evidence for Property (I) and Property (II) areas. The Property (I) area coincides with the warm base of the Amundsen crater's 35 km wide EFS, and the Property (II) area coincides with a strongly neutron-suppressed Faustini PSR. The PSRs suppression suggests evidence for heterogeneous distributions, as evidenced by the asymmetric suppression gradients bounding the PSR. Its maximum suppression location is north of its PSR's centroid, which coincides with its 73.0 K PFS base. At the PSR's poleward edge, the neutron suppression is nearly eliminated, supporting the assertion (1) that Faustini PSRs' internal WEH distribution is heterogeneous and (2) that an enhanced-WEH deposit occurs near the PFS base. Similarly note, the Figure 5(d4) profile similarly shows regolith thickening at the base of Faustini's PFS (Rubanenko et al. (2019)).

Consistent with the neutron-enhanced location at the Malapert Massifs' EFS base, Amundsen's upper EFS at 85°.6 S shows maximum temperatures exceeding 300 K and cooler ~250 K maximum temperatures near its EFS base, where the greatest neutron-enhancement occurs, Figures 5(d2)–(d4). Figures 3(d) and 4(a) show several large neutron-enhanced areas within Amundsen's basin, spanning the area between the Faustini and Amundsen PSRs. The areas have a low PSR areal density with only a few small PSRs depicted, consistent with a Property (I) area. Reflected long-wavelength radiation from the EFS, into the Amundsen's basin may again warm the surface, increase its sublimation rates, thereby creating the crater's neutron-enhanced flux.

### 3.2. Correlated Hydrogen to Maximum Temperature Distributions within the Haworth, Shoemaker, and Faustini Permanently Shadowed Region Basins

The objective of this study is to (a) quantify the WEH spatial distributions within the most detectable PSR basins within the Haworth, Shoemaker, and Faustini craters, (b) to determine if the basin's WEH is correlated to its internal maximum temperature distribution, and (c) to provide evidence that these PSRs' internal WEH distributions may be heterogeneously distributed. Jointly, the PSR correlations establish a strong line of evidence that they have internal cold-trap areas that maintain the greatest or most detectable water-ice concentrations in their surface top meter. To quantify these relationships, we evaluate these PSRs' WEH pixel distributions as a function of their latitude and the Diviner maximum temperature map (Paige et al. 2010b).

Figure 6(a) shows the latitude-dependent WEH pixel distribution and the differing north-to-south responses within the Haworth, Shoemaker, and Faustini PSRs. Figures 6(a) and 5(c4) show that Shoemaker's maximum WEH location coincides with its PSR's centroid at ~88° S (dashed blue line) and its thickened, and comparatively broader PFS. The full width of Shoemaker PSRs'

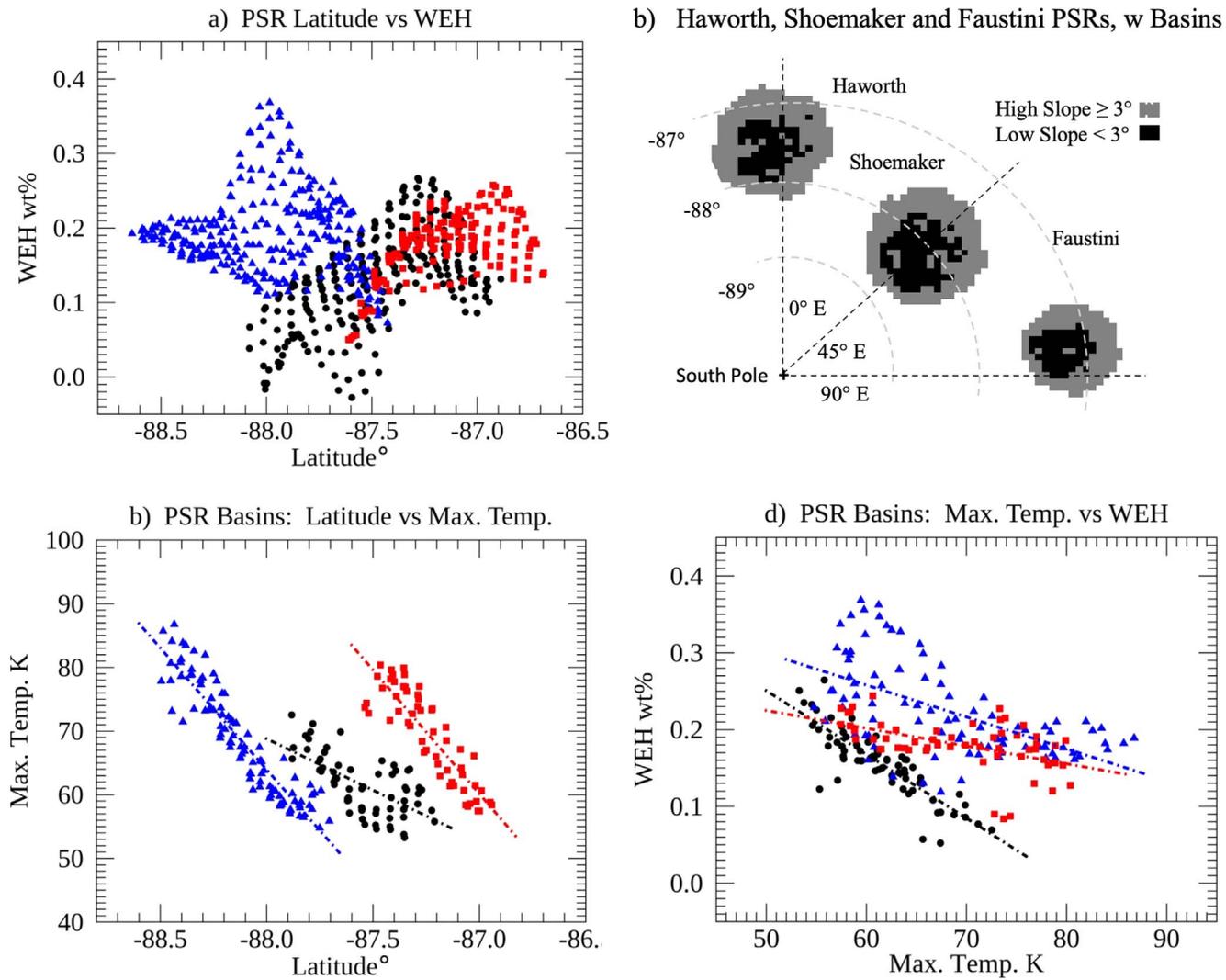

**Figure 6.** Correlation of the Haworth, Shoemaker, and Faustini PSR's maximum temperature to WEH distributions. (a) PSRs' latitude-dependent WEH pixel distributions. Dashed lines show each PSR's centroid latitude by color. (b) PSR pixels classified as high slope (gray) $\geqslant 3°$ >low slope, basin pixels (black). (c) Each PSR basin's north-to-south surface maximum temperature gradient. (d) Significant correlation of the basins' maximum temperatures vs. wt% WEH, as indicated by the negative correlation (linear fit slopes). (c) and (d) $R^2$ coefficients of determination showing the degree to which the independent variable on the x-axis predicts the basins' observations, y-axis. 0°.5 of latitude is 15.16 km (gray arrow).

**Table 2**
Haworth, Shoemaker, and Faustini Permanently Shadowed Region Studies and Statistical Test Results

| PSR | Basin: #Pixels, Area (km$^2$) | Basin: Max. Temp (K), $\mu \pm \sigma$ | Basin Temp. $\nabla$ | D Statistic | Probability $p$ | Decision |
|---|---|---|---|---|---|---|
| Haworth | 67, 268 | 61 ± 5 | −0.53 | 0.77 | 9.8E–10 | Reject |
| Shoemaker | 94, 376 | 68 ± 9 | −1.3 | 0.78 | 1.9E–13 | Reject |
| Faustini | 57, 228 | 69 ± 7 | −1.3 | 0.46 | 3.1E–03 | Reject |

**Note.** Column (1): PSR name. Column (2): PSR's basin #pixels, area. Column (3): basin's maximum temperature, mean, and standard deviation. Column (4): basin's surface thermal gradient, in units of K km$^{-1}$, where the temperature decreases south to north toward the PFS. Column (5): basin's WEH wt% versus maximum temperature, wt% WEH K$^{-1}$. Column (6): K-S D statistic. Column (7): probability the two distributions are from the same parent population. Column (8): decision; if $p < 0.01$, reject null.

WEH response is ~15 km. Basin pixel distributions for Haworth and Faustini confirm that their respective maximum WEH locations are biased to the north of their PSR centroids (dashed lines) and coincide with their PFS bases, Figures 6(a) and 5(b4) and (d4). Figure 6(b) shows the PSRs' pixels as classified into two categories of topographic slope: high slope (gray) $\geqslant 3°$ > low slope (black), which correspond to the PSRs' basin pixels.

Figure 6(c) shows the basins' north-to-south-aligned surface thermal gradients, as indicated by the negative slope of their linear fits. The thermal gradients are caused by secondary heating from the PSRs' upper EFSs. Figure 6(d) shows the correlation of each PSR basin's WEH to its maximum temperature distribution. $R^2$ coefficients of determination state the degree to which the independent variables x-axis predicts each basins' dependent variable's y-axis, maximum temperature or WEH wt%, Equation (12). Table 2 reports the metrics for each basin.

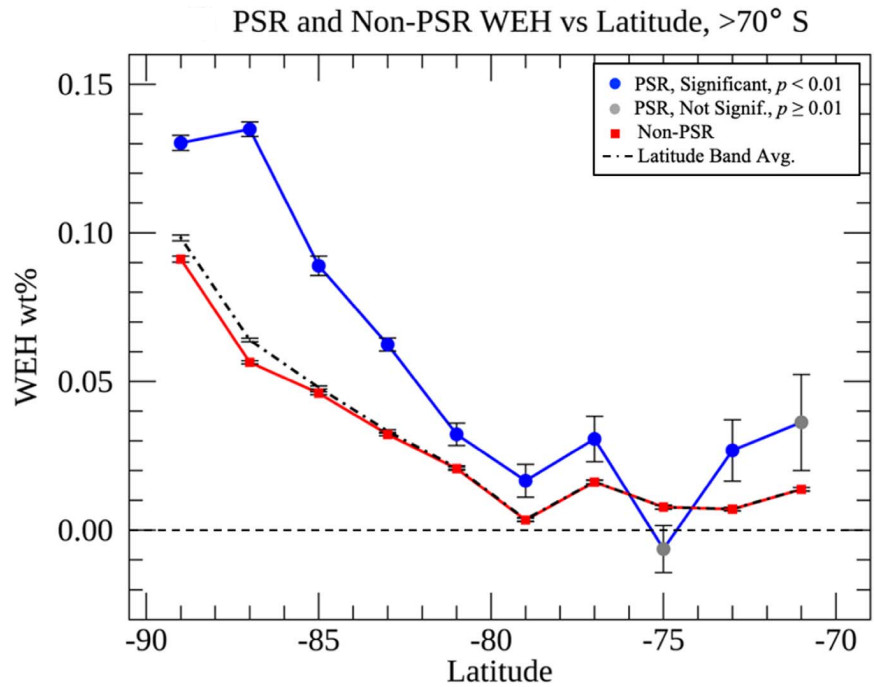

**Figure 7.** Expected PSR (blue) and non-PSR (red) WEH concentrations as a function of latitude to 70° S. Significantly contrasted PSR WEH concentrations (blue circles) vs. non-PSR WEH concentrations (red) averaged in latitude increments of 2°. The black dashed graph shows each latitude band's averaged WEH concentration. Error bars are in units of the standard error of the mean WEH. Gray circles show where a PSR's averaged WEH wt% is less than the non-PSR's averaged WEH wt% or the positive PSR WEH contrast is insignificant.

Figure 6(d) shows negative correlation slopes for all basins, as evidenced by a linear fit of the PSRs' maximum temperature to WEH distributions. Negative correlation slopes indicate that CSETN has (a) independently detected each PSR's greatest WEH concentrations within a few kilometers of their coldest basin pixels < 75 K and more importantly, (b) has detected WEH distributions that are linearly correlated to each basin's maximum temperature distribution. (c) the locations are aligned within a few km of the base of the PFS. The correlations would be substantially degraded if the PSRs' maximum temperature and maximum WEH concentrations were spatially misaligned by more than a few kilometers from their present alignments. These alignments are strongly consistent with (a) the observed polar averaged neutron suppression being greatest toward the PFS, observed in, (b) thickened PFS in midlatitude craters in Rubanenko et al. (2019), and (c) greater water concentrations found by SOFIA on midlatitude PFS slopes, relative to EFS, in Reach et al. (2023).

Statistical evidence for heterogeneous WEH distributions and cold traps within the PSR is demonstrated by classifying each basin's WEH pixel distribution into two groups based on their maximum temperature. Each basin's WEH pixels are grouped based on whether each pixel's maximum temperature is greater or less than their basin's expected maximum temperature. The K-S test was again used to determine if the two populations are drawn from the same parent population.

Table 2 results show that the low-temperature groups of each of the three independent PSRs have significantly greater WEH concentrations than their respective high-temperature groups, as indicated by their test probabilities $p$ being less than the probability $p < 0.01$ threshold. Thus, the null hypotheses are rejected, and we accept the statistical evidence the basin's hot versus cold WEH pixel distributions are significantly different. This result indicates evidence that each PSR's WEH distribution is internally heterogeneous and may indicate cold trap locations in the basins where the maximum WEH concentrations are observed. However, even though the evidence for heterogeneous WEH distributions is statistically significant, the result is not conclusive because instrumental blurring may contribute to the PSRs observed WEH distributions and gradients. Note that there is a $1\sigma$ spatial uncertainty of 5 km or 2.5 pixels related to the high-pass smoothing process, Equation (3).

### 3.3. Latitude Extent of Permanently Shadowed Region Water-equivalent Hydrogen Enhancement

Figure 7 shows PSR WEH profiles averaged relative to their respective non-PSR WEH averages, poleward of 70° S, as observed in a series of 2° latitude bins. For this study, we averaged the WEH of PSR areas toward equatorial latitudes where they are of smaller area and lesser coverage density (Mazarico et al. 2011). Property (3) from Figure 1 predicts that based on the PSRs smaller areas there should be a correlated reduction in the PSRs' expected WEH contrast relative to the non-PSR surface averaged WEH toward the equator. This study did not include LRO station-keeping zones equatorward of 83° S, which eliminates longitudes bands, $90° \pm 45°$ E and $270° \pm 45°$ E (Sanin et al. 2016).

Figure 7 plots show a contiguous set of seven independent upper-latitude bins that indicate the PSRs' averaged WEH concentrations are enhanced relative to that of their respective non-PSR averages, poleward of 77° S. A binary test of the latitude bins indicates a negligible joint probability that the seven contiguous bands could randomly show greater averaged PSR WEH concentrations relative to the non-PSR averages $(0.5)^7 = 7.8 \times 10^{-3} < 0.01$. As postulated, the poleward increase in the contrast between the PSR and non-PSR WEH profiles is correlated to the poleward increase in the expected PSR areas and the PSR areal density (Figure 1; see also Mazarico et al. 2011). At 89° S, PSR areas constitute 18.2% of that total band area, which is reduced to 3% at 81° S and 0.16% at 71° S. The PSRs' expected diameter and standard deviation decreases by latitude, as reported in 5° increments [>85° S, 80°–85° S, 75°–80° S, 70°–75° S] are [$5.2 \pm 6.2$ km, $4.1 \pm 3.4$ km, $3.5 \pm 2.3$ km, $3.2 \pm 1.5$ km].

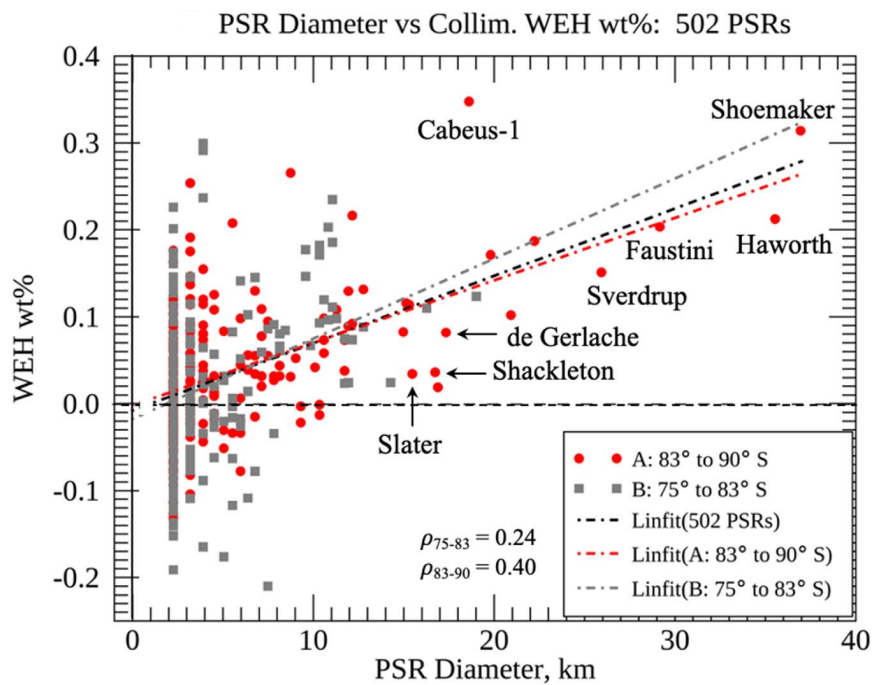

**Figure 8.** Correlated detections of PSR maximum wt% WEH vs. PSR diameter, poleward of 75° S. A positive linear correlation between the PSR diameters and their observed WEH indicates the PSRs have similar WEH distributions (except Cabeus-1). (A) 83° to 90° S (red), and (B) 75° to 83° S (gray). $\rho_{83-90}$ and $\rho_{75-83}$ are Spearman correlation coefficients for the (A) and (B) studies, respectively.

The small PSR WEH contrast equatorward of 80° S is attributed to the reduced signal-to-noise ratio of PSR areas and the region's lower coverage density. The poleward increase in the expected non-PSRs WEH is consistent with an increase in small-scale shadowing, blurring of PSRs WEH into non-PSR areas, and the expected increase in the areal density of less than 2 km diameter PSRs (Hayne et al. 2021).

### 3.4. Correlation of Permanently Shadowed Region Observed Water-equivalent Hydrogen wt% versus Diameter

Figure 8 shows the correlation of 502 PSRs' observed maximum WEH concentrations as a function of their diameter. The study's primary objective is to determine if there is evidence of instrumental blurring of the PSRs' WEH as correlated to their diameters, including PSRs poleward of 75° S. A positive correlation over that poleward latitude extent is an indication that (a) the PSRs' internally enhanced-WEH concentrations is a widespread phenomenon, (b) there is an expectation that the PSRs have similar internal WEH distributions, and (c) show further evidence of CSETN's high-spatial resolution capability. The results are realized by showing CSETN's WEH observations for the full range of PSR diameters from negligibly detected PSRs that are 2 km wide (single pixel) (left) to 37 km at Shoemaker's PSR (right).

To determine if the PSRs' observed WEH concentrations are correlated with instrumental blurring, the non-PSR latitude profile $P_{\text{non-PSR}}$ from Figure 7 is subtracted from the corresponding latitude bands of the $C_{\text{WEH}}$ map, where $C^*_{\text{WEH}} = C_{\text{WEH}} - P_{\text{non-PSR}}$. The $C^*_{\text{WEH}}$ map isolates the PSR WEH observations from their background non-PSR WEH. Two independent latitude band studies of the regions PSR population were performed: (A) high-latitude PSR distribution, poleward of 83° S (red circle) and (B) low-latitude PSR distribution, 75° to 83° S (gray squares). Image region growing was applied to the binary PSR map to derive each PSR's area and diameter, Equation (13) (Sonka et al. 1998). As in Figure 7, PSRs in LRO station-keeping zones are excluded equatorward of 83° S (B), as specified in (Sanin et al. 2017).

The low-latitude study (B) shows much greater WEH variability around its linear fit due to the region's greater statistical uncertainties, smaller diameter PSRs, and the region's lower PSR areal density relative to the high-latitude study (A). Unweighted linear fits were performed for all PSRs in each (A) and (B) study group, as well as a combined linear fit to include (A) and (B), as derived from all 502 PSRs, black (All). The unweighted fits are a conservative evaluation of the largest PSRs' WEH observations as the 1072 km² Shoemaker PSR area has the same weight in the fits as a single pixel PSR with an area $= 4$ km². The All linear model is reported as $\text{WEH}_d = -7.85 \times 10^{-3} + 7.74 \times 10^3 * d$, where $\text{WEH}_d$ is the expected WEH observation for a PSR of diameter $d$.

As predicted by Property (III) in Figure 1, CSETN's response to the smallest PSRs WEH is negligibly correlated, as evidenced by the observed high degree of variance toward km diameter (Figure 8). However, both studies indicate that the correlation increases toward the larger PSR diameters as the WEH variance around the linear fits is reduced to the right.

The All PSRs linear model shows good agreement in predicting Shoemaker's and Haworth's observed WEH wt% [0.31, 0.21],

respectively. The linear model predicts [0.28, 0.26] wt% WEH, and an average absolute difference of 0.04 wt% WEH from their respective observations, using $d = $ [37 km, 35 km]. The independent linear models from the (A) and (B) studies yield consistent predictions (A) = [0.32, 0.30] and (B) = [0.26, 0.25], respectively. Notably, the low-latitude (B) PSR linear model included no PSRs that exceed 19 km in its fit, implying those predictions are highly extrapolated results, by at least 5° of latitude and 18 km of PSR diameter. Faustini is predicted within the rounding errors by the All study, based on its $d = 29$ km, to be 0.22 wt% WEH, versus its observed 0.21 wt% WEH.

The linear correlations support the claim that Cabeus-1's elevated observation = 0.35 wt% WEH is anomalous, based on its diameter, $d = 19$ km. Its diameter-predicted observation is 0.14 wt% WEH, which implies its observed WEH concentration exceeds its prediction by a factor of 2.7. Its anomalously enhanced-WEH concentration constitutes evidence of a second, perhaps complementary WEH source that contributes to the PSRs broadly distributed WEH in the Cabeus-1 region, discussed in Section 4.

We checked the consistency of the All linear model as a predictor of the largest PSRs WEH concentrations. 100 separate evaluations were run, and for each run, a random selection of half of all the Figure 8 PSRs were obtained and linearly fit. Shoemaker's predicted WEH concentration from the linear fit was recorded for each run. From these runs, Shoemaker's predicted average WEH concentration and its standard deviation is 0.28 ± 0.03 wt% WEH, consistent with that derived from the All linear model = 0.28 wt% WEH. Shoemaker's PSR was selected because its area fully subtends CSETN's FOV, and its "true" WEH concentration *may* be observed—but we assume its maximum WEH area is uniformly enhanced (Figure 1, case P6).

The two significantly different populations have arisen during the mission as the WEH of the larger area and more detectable PSRs have shifted toward positive WEH concentrations with the accumulating coverage. With the latitude-dependent averaged WEH concentration removed, 71% of all PSR pixels poleward of 77° S indicate positive WEH concentrations. Above 83° S, the percentage of all PSR areas showing positive WEH concentrations is 73%. Positive Spearman correlation coefficients $\rho_{75-83} = 0.23$ and $\rho_{83-90} = 0.40$ reflect the expected increase in correlation as the signal-to-noise ratio represented by the PSR diameters and observation density increases from the low- to high-latitude studies, (B) to (A).

Based on the fraction of PSR area detected, a conservative correction for the PSRs' wt% WEH is proposed, as illustrated for Cabeus-1. The mixing ratio of the known footprint area $A_{\mathrm{footprint}}$ to the known PSR area $A_{\mathrm{PSRs}}$ is used. CSETN's corrected wt% WEH observation gives $C_{\mathrm{WEH\_C}} = C_{\mathrm{WEH}} * (A_{\mathrm{footprint}}/A_{\mathrm{PSR}})$. CSETN's footprint area is 707 km², $A_{\mathrm{footprint}}$, based on its 30 km diameter footprint. Cabeus-1's area is 272 km², $A_{\mathrm{PSR}}$. Their ratio generates a scaling term = 2.66. Using Shoemaker's observed $C_{\mathrm{WEH}} = 0.31$ wt% as a baseline, the model yields a corrected Cabeus-1 observation, $C_{\mathrm{WEH\_C}} = 0.81$ wt% WEH. $C_{\mathrm{WEH\_C}}$ is consistent with the LPNS image restored result, as approximated from that study's Figure 2 map $\simeq 0.9$ wt% WEH (Elphic et al. 2007). However, the utility of such a scaling approach is likely limited to the larger-area PSRs due to the lower signal-to-noise ratios of smaller-area PSRs.

Section 4 reviews the less than predicted WEH observations of Shackleton, Slater, Sverdrup, and de Gerlache shown in Figure (8) and reviews how their close proximity to neutron-enhanced areas has degraded their WEH observations.

### 3.5. Latitude Extent of Permanently Shadowed Region Areal Density Properties (I), (II), and (III)

The objectives of this study are to validate the prior observations that show the PSRs are the focal points of enhanced-WEH concentrations for the range of latitudes, poleward of 77° S. The study shows the latitude extent over which we observe all three PSR areal density Properties (I), (II), and (III), as defined in our Figure 1 model. A novel method derives the averaged WEH concentrations as a function of the radial distance inside and outside PSRs' edges, defined at 0 km. Mathematical morphology image processing operations, specifically image erosion and dilation, are used to iteratively remove or add pixel rims to a binary PSR map (Sonka et al. 1998; Mazarico et al. 2011).

The initial phase of the process erodes the PSRs' areas by iteratively stripping away the PSRs' pixel rims, thereby proceeding into Property (II) areas. Successive WEH averages within all PSRs' rims are observed from the coregistered Figure 3(a) WEH map and registered toward negative kilometer values. The smallest area, less detectable PSRs are quickly lost as each erosion eliminates 4 km from all PSR diameters, i.e., a rim of single, 2 km wide pixels. The PSRs' erosion process is exhausted when it eliminates the centers of the largest-area PSRs.

Dilation iteratively adds new PSR pixel rims, beginning at (0 km). Successive non-PSR rims are added that are of increasing distance from any PSR edge, thereby proceeding into Property (I) areas. WEH averages are reported for the series of dilated rims. The process proceeds until the available non-PSR areas are depleted. Property (I) is thereby assured as the PSR areal density is reduced with distance, to the right. Note that the distance measure is a minimum distance statement because for PSR erosions and dilations that occur in image coordinates $x$ and $y$, the distances are 2 km, but erosions and dilations that occur along diagonal image coordinates are 2 km pixels $* 2^{0.5} = 2.82$ km.

Figures 9(a) and (b) show evidence that the PSRs are the focal points for enhanced WEH, as well as evidence for all Property (I), (II), and (III) areas, indicating widespread enhancement of WEH in PSR, poleward of 77° S. Figure 9(a) plots show the high-latitude evidence spanning 83°–90° S, including (A) 83°–86° S, (B) 86° to 88° S, and (C) 88° to 90° S. The greatest averaged WEH concentrations are correlated with the largest PSR areal densities of Haworth, Shoemaker, and Faustini toward negative kilometer values, Property (II). The sharpest break in each plot's gradient is consistent with the greatest blurring at the PSR edge at 0 km, where WEH-enhanced PSR pixels are adjacent to contrasting, and anhydrous, non-PSR pixels (Figure 1). The negative WEH gradient to the right demonstrates the Property (III) instrumental blurring correlation as a decreasing PSR areal fraction is assured in CSETN's collimated FOV to the right. The plots show the least expected WEH where the PSR areal density is lowest, consistent with the neutron-enhanced areas of Figure 4.

The three poleward plots terminate in weakly WEH-enhanced non-PSR areas near 0.05 and 0.07 wt% WEH, which is likely due to Property (III) and the regions' high PSR areal density and limited non-PSR area. Note that the $G_{CL}$ smoothing of Equation (3) spatially broadens the PSRs' WEH response, with its smoothing radius defined by the line at +13 km. The plots' far (left) and (right) observations have larger error bars as fewer pixels are averaged toward each terminus.

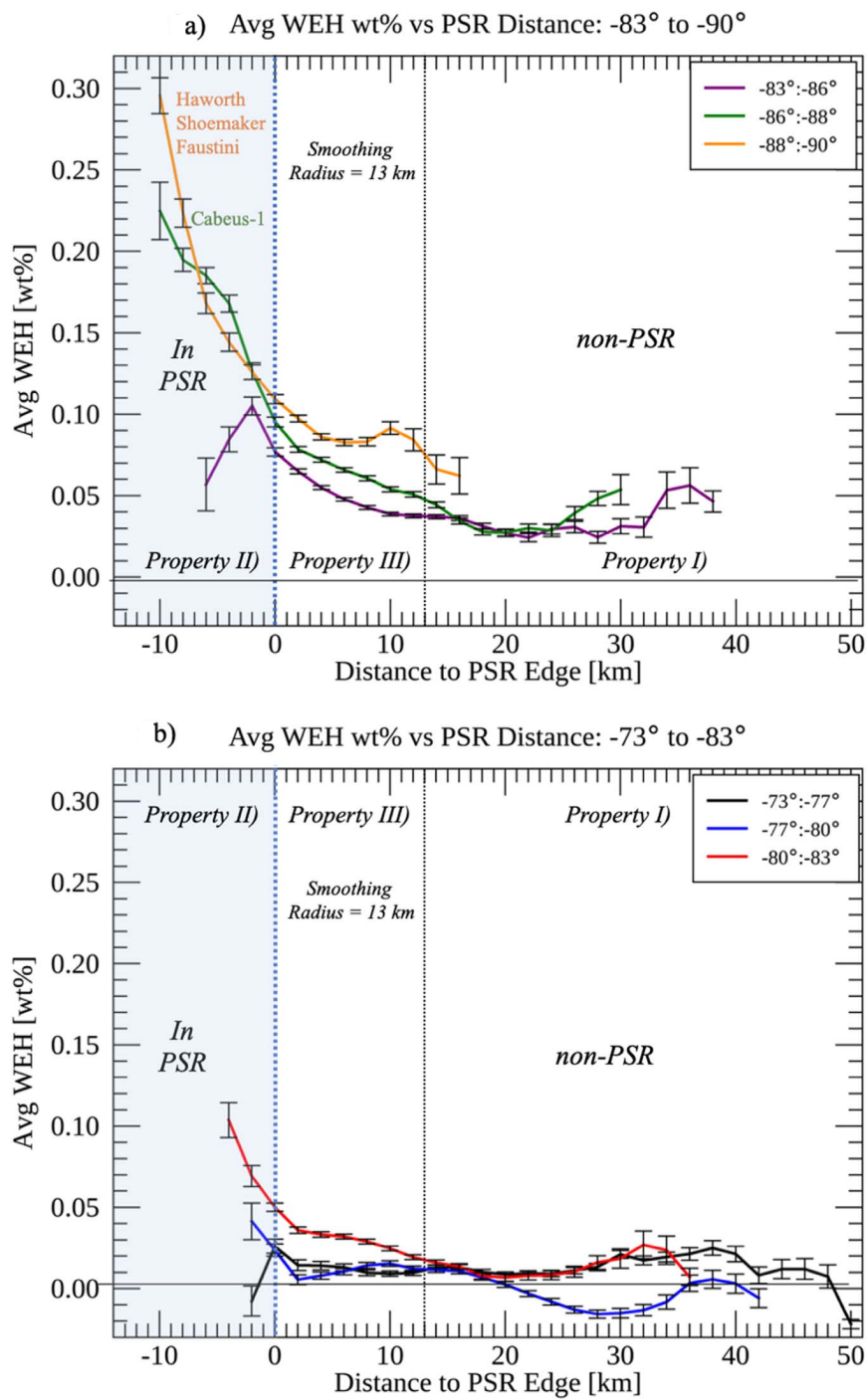

**Figure 9.** Six independent latitude band studies show evidence for enhanced PSR WEH poleward of 77° S. (a) Three independent high-latitude studies, 83° to 90° S: (A) 83° to 86° S, (B) 86° to 88° S, and (C) 88° to 90° S, (b) Three independent low-latitude bands, 73° to 83° S: (D) 73° to 77° S, (E) 77° to 80° S, and (F) 80° to 83° S. Note: the number of bins in each plot is a function of the PSR areal density and the non-PSR area in each latitude band. Error bars are in units of standard error of the mean WEH wt%.

Figure 9(b) shows the lower-latitude extent for observing all Properties (I), (II), and (III), as demonstrated for the two poleward studies, to 77° S. Three equatorward latitude band studies are shown: (D) 73° to 77° S, (E) 77° to 80° S, and (F) 80° to 83° S. The smallest-area, least detectable PSRs are shown indicated in plots (D) and (E) by the fewest negative bins are negligibly detected, and yield the predictably lowest WEH gradients. The statistical test results in Table 3 define the latitude extent. Bins of each plot were split into two groups, as those bins that are (1) closest and (2) those that are most distant from the PSRs. K-S tests of the six profiles show that the five upper-latitude tests indicated significant differences between groups (1) and (2). The (D) 73° to 77° S result indicates that the

Table 3
Kolmogorov–Smirnoff Test Results Show Instrumental Blurring Evidence in the Greater Permanently Shadowed Region Water-equivalent Hydrogen Concentrations vs. Non–Permanently Shadowed Region Water-equivalent Hydrogen Concentrations for Five of the Six Independent Latitude Bands

| Latitude Band | #Bins in (1) and (2) | D Statistic | $p < 0.01$ | Spearman $\rho$ |
|---|---|---|---|---|
| 73°S to 77°S | 15, 15 | 0.32 | 0.38 | −0.05 |
| 77°S to 80°S | 13, 13 | 0.85 | 1.2E−04 | −0.73 |
| 80°S to 83°S | 12, 12 | 0.73 | 2.4E−03 | −0.71 |
| 83°S to 86°S | 13, 13 | 0.69 | 3.0E−03 | −0.59 |
| 86°S to 88°S | 12, 12 | 0.92 | 5.4E−05 | −0.88 |
| 88°S to 90°S | 9, 8 | 0.88 | 2.2E−03 | −0.95 |

**Note.** Column (1): latitude band. Column (2): number of plot bins in groups (1) closest to PSRs and (2) most distant from PSRs. Column (3): K-S D statistic shows the maximum difference in the groups' cumulative distribution functions. Column (4): probability $p$ that the groups are from the same parent population. Bold values indicate that (1) and (2) have significantly different WEH concentrations, where $p < 0.01$. Column (5): Spearman correlation coefficient between plots' bins WEH concentrations. Negative Spearman correlation coefficients indicate negative WEH gradients from PSRs to non-PSRs.

WEH concentration difference between (1) and (2), though positive, is not significant. We also performed the same study with the mission year four version of the Figure 3(a) WEH map. We found that only the three highest-latitude studies were then significant, consistent with the convergence of these results with the ensuing coverage and the dependent reduction of statistical uncertainties.

## 4. Discussion

Figure 8 shows anomalously low WEH concentrations observed for several high-latitude, moderate-diameter PSRs, relative to their predicted WEH, identified in Figure 3(d). The PSRs and their diameters include the Shackleton [89°6 S, 129°1 E, 16.9 km], Slater [88°1 S, 114°5 E, 15.4 km], Sverdrup [88°2 S, 213°1 E, 25.6 km], and de Gerlache [88°3 S, 267°5 E, 17.3 km] craters. Figure 4(a) shows a large neutron-enhanced area, 288 km² that is ∼30 km wide (red), which coincides with the largest contiguous area of higher than average illumination, poleward of 88° S. The area coincides with Shackleton craters' outer EFS, illustrated in Mazarico et al. (2011; see Figure 6(d)). An analysis of the area's slope azimuth angle distribution indicates it is more strongly biased toward the EFS, relative to the regions expected neutron-enhanced distribution of Figure 4(b) with 4% and 36% of pixels defined as PFS and EFS, respectively. On average, the area is illuminated 25% ± 19% of the time with an averaged maximum temperature of 230 ± 60 K. At its northern edge, the area intersects Slater's PSR (Figure 4(a)). By averaging in CSETN's footprint, Slater PSRs' neutron suppression is thereby substantially reduced, resulting in its lower than expected WEH observation (Figure 8).

Shackleton crater's outer EFS is unique because it surrounds its PSR and the pole is near the crater rim. The EFS also spans ∼30 to 40 km as its elevation is reduced ∼3 to 4 km toward the Sverdrup and de Gerlach PSRs, which lie at its base. Thereby, the highly illuminated EFS generates a neutron-enhanced flux, and its near adjacency to these PSRs is similarly degrading their WEH observations. This interpretation is further substantiated by considering Shackleton PSR's weak neutron-suppression observation (Figure 5(d4)). That illustration indicates that Shackleton's WEH observation is also being substantially degraded by its adjacency to the surrounding neutron-enhanced flux from its outer EFS (Figure 8). Haworth's WEH observation is also likely degraded by its close proximity to a relatively neutron enhanced to its north, as evidenced by its Figure 5(b2) profiles, high WEH gradient.

In comparison, the comparatively enhanced-WEH observations at the largest-area PSRs (at similar latitude) are consistent with their predicted WEH, which is in part due to their greater distance from areas of neutron-enhanced flux. These larger PSRs are spatially insulated by surrounding small-area PSRs and low-illumination areas. Image restoration and machine learning methods will be needed to correct these and the other PSR WEH observations.

The PSR's WEH-to-diameter correlations of Figure 8 indicate evidence for at least two lunar WEH or other hydrogen-bearing volatile budget processes. Process #1 is likely a global phenomenon evidenced by the consistency of the two linear correlations, from the two independent PSR WEH to diameter studies . Process #1 implies the WEHs are randomly and similarly distributed to most, if not all, PSRs poleward of 77° S. The linear correlations would not be strongly aligned if the differing detectable areas of the two PSR distributions did not yield an area dependent, fractional detection of the PSRs WEH, Figure 1. Though not evaluated in this study, we presume that Process #1 should be operating in the northern hemisphere as well. Process #2 yields anomalously elevated WEH near the Cabeus-1 PSR, discussed below.

The Figure 8 observation that most PSRs are relatively and similarly WEH enhanced relative to their background WEH strongly indicates an ongoing process that introduces hydrogen to the surface, most likely as $H_2O$, but possibly to include $H^+$, $OH^-$, $H_2$, or other hydrogen-bearing species. CSETN cannot differentiate their species. However, their sources may include outgassing from the lunar interior, solar-wind deposition of protons ($H^+$) to the surface, or hydrogen-volatile species created after proton bonding with regolith silicates (Arnold 1979; Crider & Vondrak 2000; Starukhina 2001, 2006; Crotts & Hummels 2009; Pieters et al. 2009; Sunshine et al. 2009).

The Figure 8 observation supports evidence that the volatiles are part of a random migration process that uniformly distributes volatiles to the surface. Only in the stable and cryogenic PSRs, and to a greater extent their cold traps, are surface residence times long enough for water ice to accumulate near the surface (Watson et al. 1961). Water migration may arise from small-scale impactors that eject molecules and perhaps other hydrogen-bearing species in random directions and velocities, which would be deposited uniformly on the surface (Moores 2016). Randomized surface migration of hydrogen volatiles, i.e., $OH^-$, may also arise from diurnal surface temperature variations (Pieters et al. 2009). In this case, high temperatures approaching noon time are thought to induce $OH^-$ to migrate from warm locations toward greater lifetimes approaching the terminators, the poles, PFS, and the PSRs. Such a migration process is indicated in both infrared and ultraviolet observations (Pieters et al. 2009; Li & Milliken 2017; Hendrix et al. 2019). The balance of expectedly lower surface residence lifetimes and high sublimation rates, coupled with low arrival rates should rapidly deplete water and hydroxyl concentrations for even episodically illuminated non-PSR surfaces, as evidenced in the detection of water ice/gas plumes emanating from illuminated surfaces and not PSR, thought to be sublimation events, observed by near-infrared observations of the Spectral Profiler on board the Japanese Selenographic and Engineering Explorer (SELENE/ Kaguya; Ohtake et al. 2024).

The primordial impacts of comets and meteors may have introduced large water-ice concentrations at their surface impact locations. However, over billions of years, any residual water ice would likely have been randomly redistributed in subsequent impacts (Arnold 1979; Prem et al. 2015). Impact modeling studies indicate that subsequent gardening, excavation, and reburial by secondary impacts would leave very little of this water within the surface top meter or within the PSRs, which would be detectable by neutron methods (Ong et al. 2010; Costello et al. 2021).

We note that Cabeus-1's anomalously enhanced-WEH observation may be partially explained by its >20 km distance from neutron-enhanced areas (Figure 4(a)). Cabeus-1 PSR has a surrounding distribution of very low-illumination surfaces and smaller-area PSRs. These areas isolate Cabeus-1's neutron suppression from being otherwise degraded by CSETN's footprint averaging with neutron-enhanced areas (Figure 4(a)). Its surrounding PSRs' WEH may also bias its anomalous maximum WEH location, as well as act to broaden to its neutron suppression (Figures 5(a1)–(a2)).

The coincidence of the maximum WEH locations to the basins' cold PFS bases observed in Figures 5 and 6 supports the topographic evidence for water ice (Rubanenko et al. 2019). Regolith thickening at the base of the PFS is illustrated for the Haworth, Shoemaker, and Faustini PSRs, Figures 5(b4), (c4), and (d4), respectively. The Rubanenko et al. (2019) study found that the depth-to-diameter ratios of simple craters, of less than 15 km diameter, on the Moon and Mercury were reduced by regolith thickening within their basins by as much as 50 m, poleward of 75° S. The greatest thickening occurs toward the craters' PFS base. The study concluded that the regolith thickening is due to water-ice deposits accumulating toward the base of the craters' PFS as an admixture of water ice and regolith.

The Figures 4 and 5 observations are highly relevant to the coming lunar surface missions that plan to assay PSR water-ice deposits, e.g., Artemis and the planned Volatiles Investigating Polar Exploration Rover (Colaprete 2021). Figure 4 maps where there is a lower expectation of water ice deposits, i.e., that they are smaller, less frequent, or of lower WEH concentration, or they are buried. Independent results for the three largest-area PSRs show the greatest WEH concentrations coincide with the cold, <75 K base of their crater's broadened and show relatively thickened PFS bases. The alignments indicate either that the deposits have the highest top-meter WEH concentrations, or these deposits have a relatively larger and more detectable surface area/volume, or they are relatively closer to the surface than deposits located in warmer locations. The scenarios all imply that areas near the base of the PFS maintain the most efficiently accessible water-ice deposits (less overburden, highest WEH concentration).

Conversely, the Figures 5(a)–(d) evidence of lesser WEH concentrations toward the basins' warmer EFS may also be attributed to less WEH in the top meter or a gradually thickening anhydrous layer overlying a deposit. The anhydrous layer is thereby thinnest toward the PFS and thickest toward the EFS. The thicker EFS side of the anhydrous layer would effectively reduce the deposit's observed WEH observation. Both scenarios likely require manipulating a greater volume of anhydrous overburden to access subsurface deposits toward the EFS, as compared to the PFS. Non-PSR results from the Figure 4(a)–(b) neutron-enhanced areas support this claim, as EFS conditions of less shadow and warmer temperatures are less likely to contain cold traps or larger or more efficiently extractable water deposits, relative to the PFS, as indicated in Figure 4(a).

The possibility that water molecules can migrate above the surface on ballistic trajectories from meteor impacts may explain how the highest WEH concentrations occur toward the PFS base. A catchment may be formed by the relatively greater amounts of shaded PFS area above the PSR basins, as compared to their respective EFS, observed in Figures 5(b4)–(d4). The catchment model entails the deposition and high residence times of migrating water molecules maintained on the PFS. Water deposited toward the warmer EFS would have a lower expected residence time and a lower likelihood of migrating and accumulating at the EFS base. Mass wasting or ejecta from subsequent small meteorite impacts may drag any entrained volatiles down slope and focus them toward the PFS base by the craters' concentric internal slopes. The process would create deposits of admixed water ice and regolith near the PFS base. The catchment process would be further enabled if the volatiles' dominant migration direction is poleward, as the process would create higher deposition rates on the craters' PFS relative to their EFS (Moores 2016).

Our results suggest that the broad polar epithermal neutron suppression detected by (Feldman et al. 1998; Mitrofanov et al. 2010a) is explained by CSETN's detection of Property (I), (II), and (III) areas poleward of 77° S. Mazarico et al. (2011) showed that the expected PSR areas and their areal density increase toward the pole, so the PSRs' neutron-suppression observations are increasingly convolved there. Figure 7 shows the correlated poleward increase in the PSR WEH contrast versus the non-PSR WEHs. Figures 4 and 9(a)–(b) also support this claim in their correlated detection of WEH as a function of PSR areal density.

Localized areas of neutron-enhanced flux were first described in craters from LPNS observations (Feldman et al. 2001). A subsequent evaluation of 2215 crater basins quantified the enhancement and termed the effect "neutron beaming," but its source has not yet been identified (Eke et al. 2015). From the crater basins, that study found an expected 1% enhancement in their epithermal neutron-emission flux. Following our Figure 4(a)–(b) results, we postulate that the craters' observed neutron-enhanced flux coincides with the neutron-enhanced flux areas and their EFS biased properties identified in this study. A cross validation of the most neutron-enhanced locations and the evaluated crater distribution is required to check this claim.

CSETN has detected indirect evidence that enhanced-WEH concentrations exist in PSRs smaller than the baseline scale of 2 km pixels evaluated in this study. The present study found the lower-latitude extent for enhanced WEH observed by CSETN in PSR to be poleward of 77° S. Equatorward of that latitude the available PSRs are simply too few, too small, too warm, and are less detectable. However, McClanahan et al. (2015), Figure 7(b), showed that the latitude range for detecting surface WEH may be expanded toward at least ∼65° S, by showing the poleward regions WEH are consistently enhanced toward PFS relative to EFS. WEH-enhanced PFS are likely detectable because each pixel represents, on average, a greater total area of smaller PSR area, ≪2 km pixels, than their respective EFS pixels, as demonstrated in Figures 4(a)–(b).

Further, that study showed that the spatial scale of the PFS slopes was a factor in the observed geater neutron suppression of large vs. small-area PFS, to at least 65° S, as predicted by Figure 1. Over the same latitude range there was no consistent neutron suppression contrast between large and small-area EFS, implying their neutron suppression does not contrast from their surroundings to at least 65° S. Thereby, the latitude extent of evidence for water ice on PFS is extended to be consistent with the definitive water evidence on PFS observed by SOFIA at Clavius crater at 58° S. The joint results would suggest that the water volumes on PFS are far greater than the top few microns that are observable by SOFIA. A joint study is required to better understand these results.

Several potential explanations exist as to why we observe the neutron-enhanced areas near the pole: (1) The poleward 5° of latitude are over 450 km from the anhydrous latitude band location, so there may be a geochemically induced difference in counting rates, Equation (2). (2) The neutron-enhanced areas have a lower WEH concentration than that count rate average taken at the anhydrous latitude band, as suggested by their warm EFS bias, Figures 4(a)–(b), and as evidenced for Shackleton craters' outer EFS. (3) The uncollimated scaling term of 0.5 defined in Equation (5) may need to be decreased. A smaller term, less than 0.5, reduces the uncollimated neutron suppression and adds suppression (WEH) to the polar PSRs. Figures 5(a)–(d). (4) The polar latitudes and the anhydrous latitude band had very different CSETN observing conditions: from differing times of the solar cycle, varying altitude distributions, varying orbital trajectories, varying orbital inclination angles, and were observed with differing detector combinations, which could yield different counting rates for the two regions.

In any event, a possible correction to the anhydrous location or its counting rate is expected to be small, on the order of +0.021. Note that we evaluated this possibility by shifting the anhydrous suppression threshold from 1.0 to 1.02. A rerun of the study in Figures 4(a)–(b) showed that the resulting area of neutron-enhanced pixels is reduced by 58% and its slope azimuth distribution profile has an even greater bias toward the EFS, relative to that of the Figure 4(b) result. The result indicates the adjustment yields a better location and threshold for the anhydrous counting rate and its definition is justified by these physical properties.

We demonstrated in Section 3.1.1 that the spatial bandpass filter is not enhancing the PSRs' neutron suppression and manually validated our processing pipeline. The bandpass filter design and its application are unbiased as it smooths with spatially symmetric kernels, Equations (3)–(6). Further, the lunar topography, PSR locations, slopes, and temperature distributions used to select and test the epithermal neutron populations are independently defined by volcanic and impact processes. Hence, any subsequent WEH map analysis, is also unbiased.

## 5. Conclusions

We report that the Moon's south pole hydrogen distribution, most likely in the form of water ice, is widely distributed within its PSRs, poleward of 77° S. We modeled and demonstrated a comprehensive line of geophysical and geochemical evidence to explain, using observations of the region's epithermal neutron flux, the spatial distributions of water ice within the surface top meter. The observations were made by the CSETN, which is part of the LEND operating on board the LRO.

Our results are highly relevant to the planned on-surface robotic and human investigations of the lunar surface by NASA's Artemis and Commercial Lunar Payload Services missions. The study quantifies the geophysical conditions and provided maps detailing where the highest and lowest water concentrations are expected considering the surface top meter. Based on our independent studies of the Haworth, Shoemaker, and Faustini PSRs, the greatest hydrogen concentration locations are demonstrated to be coincident with the coldest surfaces and near the base of the PSRs' PFS. A map of relatively anhydrous areas defined by relatively neutron-enhanced flux demonstrates they are biased toward warm and illuminated EFS. These results are also highly relevant to research on lunar and solar system volatiles, as well as solar system origins research.

Our study predicted, emulated, and demonstrated results that are strongly consistent with the theoretical results of Watson et al. (1961). That study found that the PSRs' cryogenic surfaces sequester water-ice accumulations near the surface, likely due to their extreme vacuum and their low sublimation rates. We used instrumental blurring properties to demonstrate that the PSRs' observed hydrogen distributions are correlated to their areal density. The correlation is caused by the mixing ratio of relatively neutron-enhanced non-PSR areas and the PSRs similarly neutron-suppressed areas that are observed within CSETN's collimated 30 km diameter footprint. Two independent latitude band studies identified positive and consistent linear correlations between the PSRs' observed hydrogen concentrations and their diameters. The correlation from their combined studies indicates an expectation that the PSRs have similar hydrogen concentrations at $0.28 \pm 0.03$ wt% WEH (lower bounds). The correlation was made from the WEH observations of 502 PSRs with diameters ranging from 2 to 37 km, in latitudes ranging from 75° to 90° S.

Nearly all PSRs have detectable areas that are smaller than CSETN's footprint area, so these PSRs can only be fractionally detected. The statement implies that all PSRs must be lower-bounds WEH observations, with the possible exception of Haworth, Shoemaker, and Faustini. The statement assumes that the PSRs' hydrogen is uniformly distributed across their area. However, we demonstrated evidence for the greatest hydrogen concentrations in the PSR occurring at their cold traps, as subsets of the PSR areas that maintain maximum temperatures below 75 K and are located near the PSRs' PFS. The evidence of substantially smaller cold-trap areas imply that most of the the PSRs' hydrogen observations are underestimated.

We demonstrated that the epithermal neutron-emission flux is correlated to the PSRs' areal density, poleward of 82° S. A map of neutron-enhanced regions depicts locations where the lowest hydrogen concentrations are expected. Neutron-enhanced areas have lower than expected PSR areal densities and lower than expected PSR areas. Neutron-enhanced areas are significantly biased toward EFS surfaces, which have smaller shadows and higher temperatures than neutron-suppressed areas.

In contrast CSETN observes the most neutron-suppressed areas are significantly biased toward cold PFSs. Evidence for the greatest hydrogen concentrations are coincident with the highest PSR areal densities that occur at the Cabeus-1, Haworth, Shoemaker, and Faustini craters. High-spatial-resolution studies of these PSR basin's pixel distributions indicate their internal WEH distributions are significantly correlated to their maximum temperatures, with their maximum internal WEH concentrations occurring near the base of their PFSs at locations that are below 75 K. Our hypotheses are unified by demonstrating that the transition between the neutron-enhanced and neutron-suppressed observations is explained by the instrumental blurring and the correlation of similarly neutron-suppressed (WEH-enhanced) PSR areas.

The results showed that Cabeus-1 has an anomalously enhanced hydrogen concentration of 0.39 wt% WEH for its area and an anomalously located maximum hydrogen concentration [84°.35 S, 331°.78 E]; neither result is consistent with those measures predicted or observed for the other PSR WEH observations. These observations suggest that a second hydrogen budget process is augmenting the region's hydrogen concentrations. Alternatively, Cabeus-1's hydrogen observation may be augmented by hydrogen concentrations in its surrounding cluster of small PSRs and low-illumination surface.

We developed a specialized bandpass filter to subtract CSETN's detection of uncollimated neutrons. PSR profile studies of Cabeus-1, Haworth, Shoemaker, and Faustini validated its correct operation and our processing pipeline. CSETN's high-spatial-resolution capabilities were demonstrated in its collimated WEH map and in profile studies that all show the narrow width of its response to the larger neutron-suppressed and neutron-enhanced areas with their contrasting geophysical contexts. We showed that CSETN detects the highest hydrogen concentrations coinciding with the coldest PSR basin locations, occurring at the base of their PFS. We also demonstrated that CSETN proportionally detects PSR areas, which establishes evidence that they have similar hydrogen distributions. Thereby, we consider claims that assert CSETN's FOV performance is no better than an uncollimated neutron sensor and that its collimated counting rate is negligible can be dismissed (Eke et al. 2012; Teodoro et al. 2014).

Our objective in this study was to develop a comprehensive understanding of the lunar epithermal neutron-emission flux by characterizing it in the context of volatile-related geophysical observations and by demonstrating the expected degradation from instrumental blurring. We mapped at 2 km pixel resolution, much higher than CSETN's detection capability to demonstrate instrumental blurring in the correlated detection across a range of PSR areas, including to show the expected detection failure for the smaller PSRs. Single pixels and the smaller PSRs have very low statistical significance and can only be evaluated in bulk averages, and in the greater context of the correlated observations. The study did not factor statistical uncertainties that are important for understanding the statistical significance of specific areas. Note that there are moderately neutron-suppressed areas, especially toward the equatorial latitudes where the statistical uncertainties are higher, and which are not well aligned with PSRs and may be false positives. Subsequent polar studies of CSETNs observations should consider the geophysical context in their analysis.

Future work will require machine learning methods to understand the hydrogen concentrations and detection significance of specific areas. The methods will enable revision of the background estimate, identify specific cold-trap areas and their contributions to NSR, and to integrate observation counts, e.g., Cabeus-1.

LRO's orbital inclination now precludes the nadir-pointing coverage of the larger-area high-latitude PSRs, which concludes those observing campaigns. The LRO planning ephemeris shows that its orbital inclination will continue to degrade in the coming years and stabilize toward 83° S by 2026. In that span, enhanced coverage will ensue in the latitude band between 83° and 85°, which will improve the coverage and statistics of important lower-latitude PSRs at the Cabeus-1, Amundsen, Idel' son, Weichert, Nobile, and Malapert regions.

## Acknowledgments

We thank the Lunar Reconnaissance Orbiter (LRO) project science team for its substantial support of this effort and their ongoing contributions to LRO and lunar science. The effort was sponsored in part by the NASA Goddard Space Flight Centers Artificial Intelligence Working Group as well as NASA grant award number 80GSFC21M0002. We acknowledge the use of imagery from Lunar Quickmap (https://quickmap.lroc.asu.edu), a collaboration between NASA, Arizona State University, and Applied Coherent Technology Corporation. This study could not have been made without the provision of high-spatial-resolution maps provided by the LOLA and Diviner instrument teams. We deeply appreciate the ongoing contributions of the NASA Planetary Data System (PDS) for their continued curation and maintenance of these critically essential lunar archives.

## Data Availability

This article's data and software repository can be found on Zenodo: doi:10.5281/zenodo.10027812.

# Appendix
# Secondary Studies, Products and Materials

*A.1. Figure A1 South Pole Maps from LRO Instruments to 80°S*

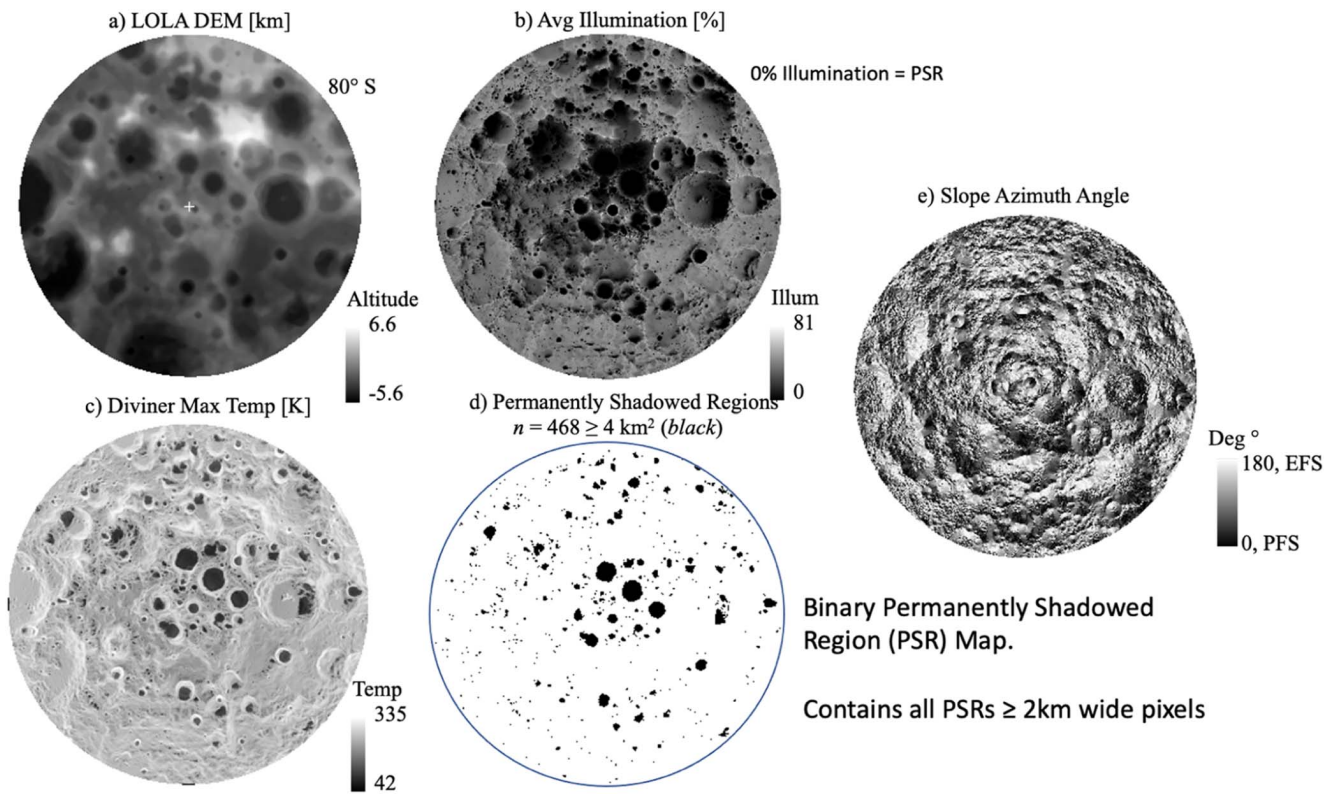

**Figure A1.** LRO coregistered maps used in this study. (a) LOLA digital elevation model as deviation from the 1737.4 km mean lunar radius (Smith et al. 2010). (b) Averaged illumination map derived from the lunar topography in (a) as averaged over several lunar precessions using lunar and solar ephemerides (Mazarico et al. 2011). (c) Diviner maximum temperature map derived from observing surface maximum temperatures during the first years of the Diviner mission (Paige et al. 2010b). (d) Binary PSR map derived from pixels that have 0% averaged illumination from map (b). Map pixels: 0 = PSRs and 1 = non-PSRs. (e) Slope azimuth angle map derived from the LOLA topography, where 0° is a PFS, 180° is an EFS, and 90° is an east- or west-facing slope.

*A.2. Figure A2: LEND Coverage Distribution Profiles by Mission time and South Pole Latitude*

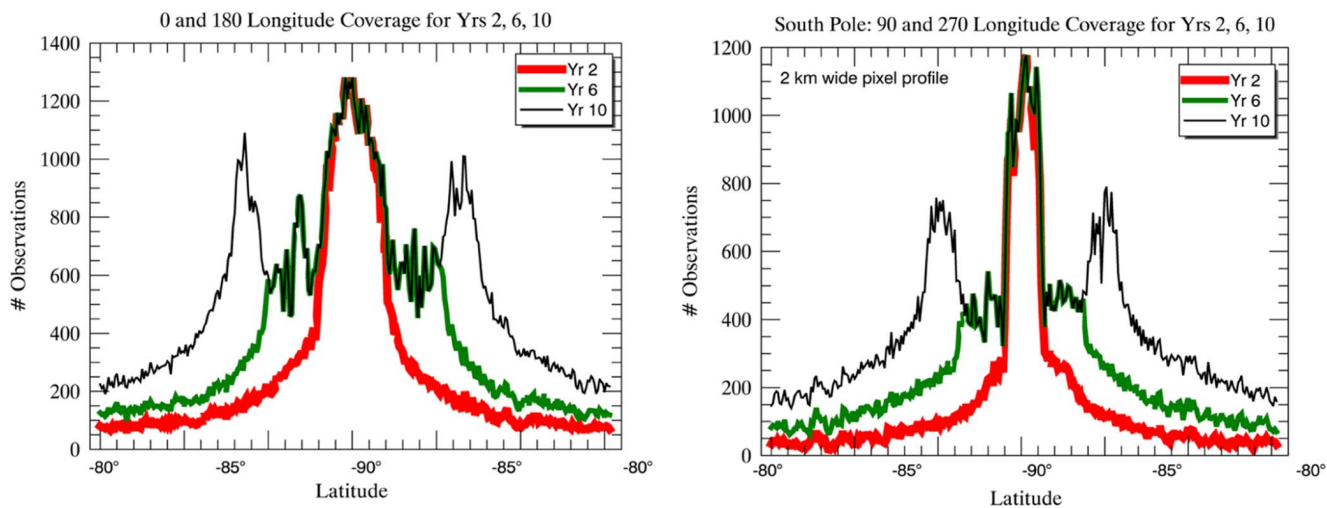

**Figure A2.** Time-dependent pixel observation counts by latitude in a pixel profile through the south pole in a 2 km wide band for mission years 2, 6, and 10. Profiles show CSETN's south pole latitude distribution of coverage as it evolved through the mission. Red shows the first two mission years and a high density of polar coverage, mostly occurring poleward of 88° S. Note that Shoemaker had a significant amount of its accumulated coverage during this time. The orbital inclination slowly degraded after mission year 2, which shows the relatively increased coverage accumulations toward lower latitudes (green) for year 6 and (black) year 10, with little additional coverage at high latitudes after year 2. Relatively reduced coverage occurs in the 90° and 270° longitude profiles (right plot) due to early mission station keeping in those longitudes.

*A.3. Figure A3 LRO Apsis History 2009–2021*

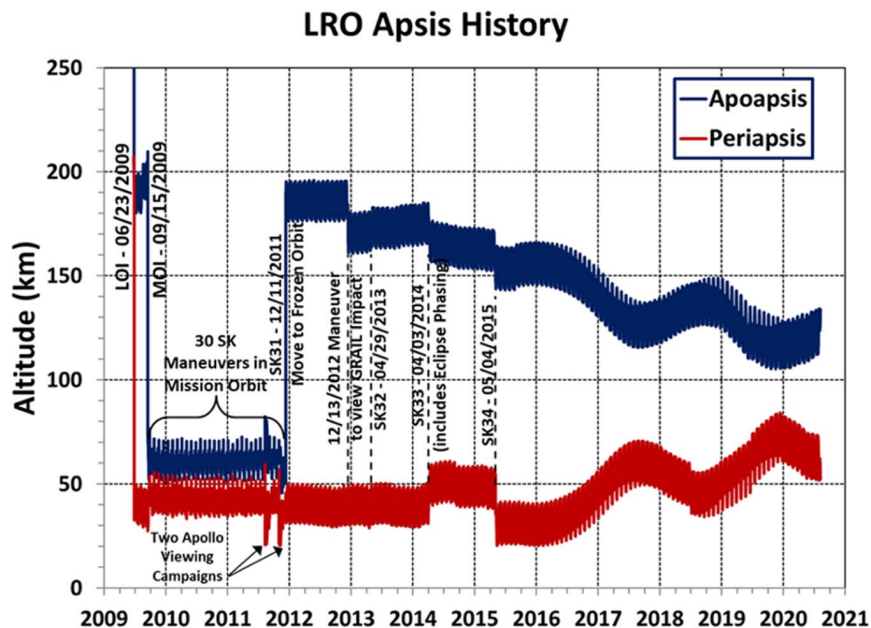

**Figure A3.** LRO Apsis history shows the variation in orbital altitude for the north and south poles for LRO's mission phases. The commissioning phase had an elliptical orbit near 40 km in the south and 190 km in the north, spanning 2009 July 2 to 2009 September 14. The science mapping phase had a nearly circular orbital altitude, near 50 km, which spanned 2009 September 15 to 2011 December 11. Several LRO extended mission phases all had elliptical orbits with evolving polar altitudes spanning 2011 December 12 to the present. The south pole periapsis increased from 2015 to 2021. In 2022 and 2023, LRO's south pole apsis continued to process toward a higher average altitude.

*A.4. Figure A4: The LEND, Instruments and Sources*

License to use LEND instrument figure.
Official LEND description: Mitrofanov et al. (2010a).
Published LEND calibration and analysis: Boynton et al. (2012), Litvak et al. (2012a, 2012b, 2016), Livengood et al. (2018), Mitrofanov et al. (2008), and Sanin et al. (2012, 2017).

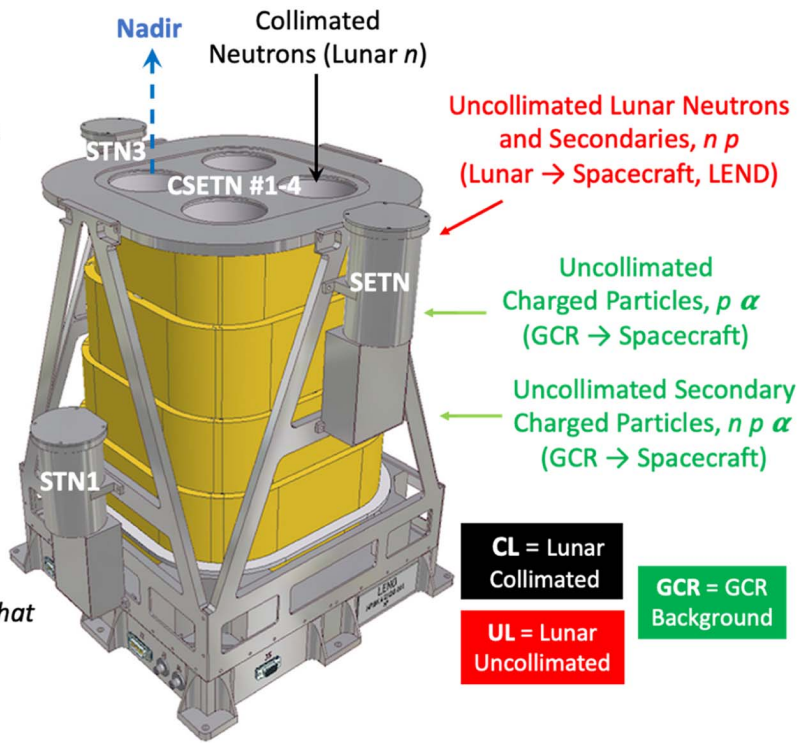

**Figure A4.** The LEND, its suite of detectors, and its contributing sources. Shown in yellow is the collimator assembly for the four $^3$He detectors of the CSETN, with their circular apertures at the top. The collimator is made of neutron-absorbing materials B$^{10}$ and polyethylene to discriminate against the detection of neutrons originating from outside CSETN's collimated FOV. The uncollimated Sensor for EpiThermal Neutrons (SETN), also a $^3$He detector is external to the collimator (front, right). Three Sensors for Thermal Neutrons (STN1-3) are $^3$He detectors that surround the collimator. CL neutrons are detected after coming through the detector apertures from their lunar source and are absorbed in the $^3$He detectors. CL neutrons detected from this configuration are detected at high spatial resolution. UCL neutrons start as high-energy epithermal or fast lunar neutrons that can either be directly detected after passing through the collimator materials, or they are detected as epithermal energy neutrons after their energies are attenuated after scattering after interacting with the collimator or spacecraft materials. UCL neutrons carry a signature of neutron suppression at the poles, which we subtract using our spatial bandpass filter. Neutron interactions with the collimator and spacecraft can also create a dependent flux of charged particles that may be detected. UCL neutrons are low-spatial-resolution detections. GCRs interact with the collimator materials and are detected. The flux is independent of the lunar surface corrected and is subtracted in ground calibration.

*A.5. Figure A4: CSETN and SETN: GEANT4 Modeled Energy Distributions*

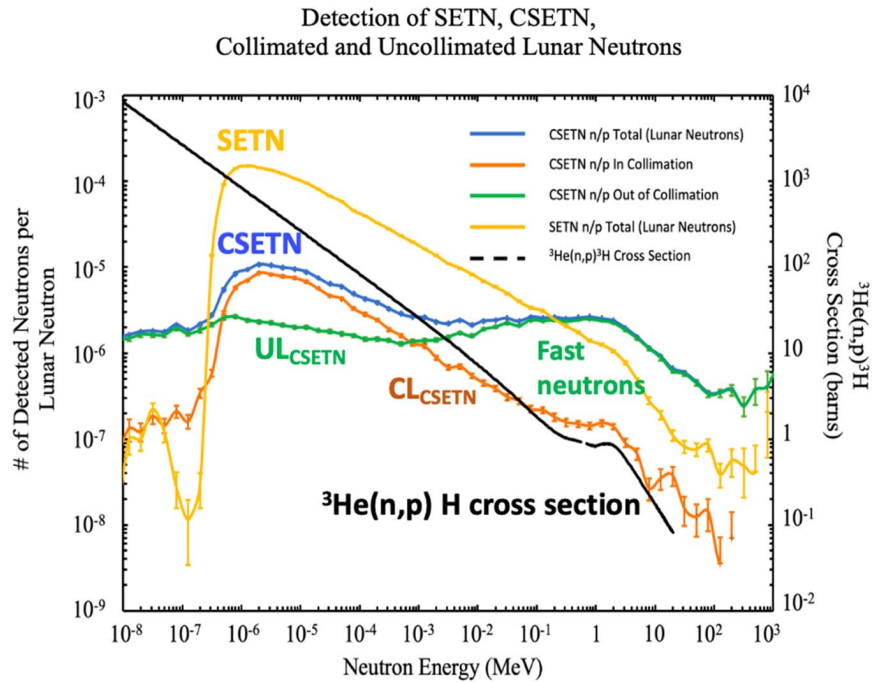

**Figure A5.** Neutron transport GEANT4 modeled energy distributions as detected by SETN and CSETN. CSETN's total lunar neutrons are in blue along with the energy distributions of UCL (green) and CL neutrons (orange). CSETN's UCL neutron component (green) originates as high-energy epithermal and fast neutrons whose energies are down scattered to epithermal energies after interacting with the collimator or spacecraft materials. CSETN's uncollimated neutrons are of low spatial resolution and contain a signature of hydrogen, as neutron suppression. CL (blue), UCL (magenta), CSETN total (red), and neutron-scattering cross sections (black). Note that CSETN's collimated energy distribution (orange) is of the same shape as SETN's, indicating their $^3$He detectors detect the same neutron energy distribution, but CSETN has a lower counting rate.

*A.6. Figures A6(a), (b): CSETN, SETN: GEANT4 Modeled Angular Response to Lunar Neutrons*

In Figure A6(a), the CSETN WEH study indicates that both collimated and uncollimated neutrons are sensitive to varying regolith hydrogen concentrations that are uniformly distributed in the surface top meter. A reduction in the observed flux indicates neutron suppression as hydrogen concentrations increase relative to the anhydrous case, WEH = 0% (black), where WEH = 0.1% is in blue and WEH = 1% is in red. Note that the WEH-dependent neutron suppression spans both collimated and uncollimated contributions. This implies that the UCL neutron energy distribution is also sensitive to regolith WEH concentration. The result implies that $U_{CL}$ neutrons will also cause neutron suppression, evidenced in the map in Figure 3(c) and the PSR profiles in Figures 5(a)–(d).

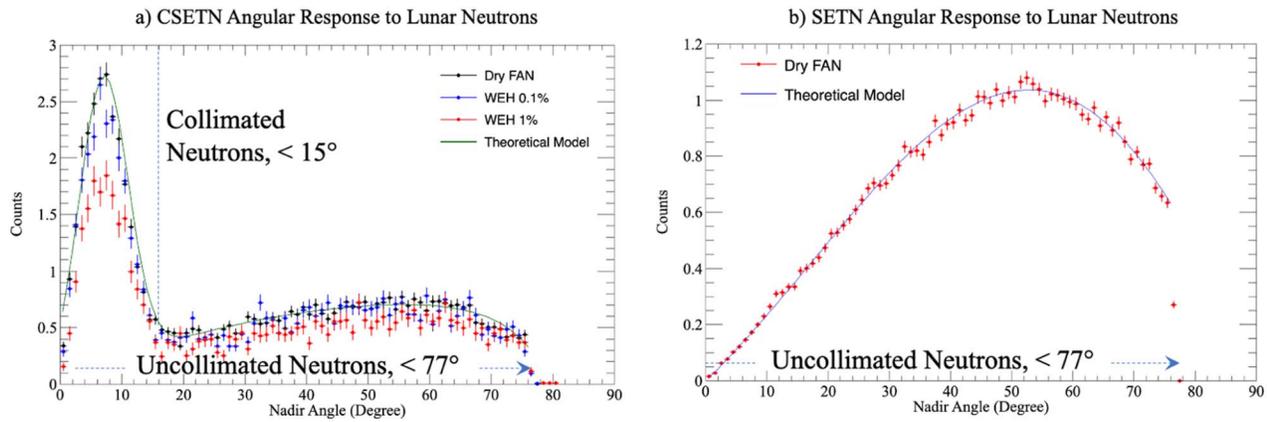

**Figure A6.** CSETN (left) and SETN (right) angular response to lunar neutrons as modeled in GEANT4, from a 50 km altitude. (a) CSETN's angular response to the lunar neutron-emission flux. The lunar neutron-emission flux is dependent on the water concentration being uniformly distributed within the regolith (WEH wt%). The plot is integrated for 360° around the nadir orientation, angle = 0°. CL neutrons are exclusively detected within 15° of the instrument boresight. At greater degrees off nadir, lunar neutrons must interact with and be scattered by, or pass through, the collimator materials to be detected. Nadir areas are small, so they have relatively few detected counts. As the angle increases, the counts increase with area until 5°.12 from nadir. The counts decrease from 5°.12 to 15° because the $^3$He detector areas are increasingly obscured by the collimator materials, which reduce the detected counts. A small fraction of the neutrons that arrive within 15° of nadir are technically uncollimated because they interact with the collimator faceplate, which are scattered and detected. The lunar limb is encountered at 77°, which terminates the angular range for lunar neutrons. (b) SETN's angular response for incidence angles in degrees relative to the nadir-pointing vector (right plot).

### A.7. Figures S7(a), (b): CSETN and SETN: GEANT4 Modeled Fields of View and Footprint

*Discussion on GEANT4 modeling and sources of error.* We developed our independent GEANT4 models of the LEND detectors to independently investigate the LEND team's published claims of CSETN's detection of PSRs and the published counterclaims of its poor performance. An initial report of the study and its objectives was documented in an LPSC abstract by Su et al. (2018). LEND and CSETN instrument geometry and compositional information are considered proprietary and are not available for dissemination with this paper.

The LEND instrument PI is Dr. Igor Mitrofanov at Space Research Institute, Russian Academy of Science (IKI), Moscow Russia.

Our models were developed in GEANT4 and used LEND composition and geometry information from the IKI team, as well as from published descriptions. The IKI team originally developed their models in the Monte Carlo N-particle code, but we understand that IKI has in recent years shifted to GEANT4. Our WEH conversions were found to be in good agreement with Sanin et al. (2017) as below and were adopted.

GEANT4 modeling of neutron capture events yields a low uncertainty for CSETN's total, CL, and UL components. Their neutron captures are on the order of 15,000, 6000, and 9000 counts, respectively using $5 \times 10^9$ incident neutrons of lunar energy spectra, which factors the emission flux from lunar soils with a range of chemical composition and hydrogen concentration. The captures are translated to an uncertainty for LEND observations below $10^{-5}$ cps.

Lunar neutron spallation production is calculated using charged-particle energy spectra based on the Usoskin et al. (2011) model to bombard lunar soils of different chemical compositions for calculating neutrons emitted from the lunar surface. In each case, we used $10^6$ GCR protons and alpha particles as the input to simulate the spallation neutron production process that in average produces $3 \times 10^7$ neutrons. This figure can be translated into an uncertainty of lunar neutron flux in the order of $10^{-10}$ cm$^{-2}$ s$^{-1}$. Thus, the uncertainty of the lunar neutron flux can be totally neglected. A similar result is found for the LRO spacecraft-induced events, where the uncertainty of GCR components is in the order of $10^{-4}$ cps for the LEND detection rates. So, the GCR precision error is neglected in this study.

Our independent GEANT4 neutron modeling analysis of the LEND/CSETN system shows CSETN's collimated count rate is 0.93 counts s$^{-1}$, which is well within the $1.0 \pm 0.1$ counts s$^{-1}$ total count rate (four detectors), and its uncertainty reported in Litvak et al. (2016), and used in Sanin et al. (2017). Though our modeling would indicate more neutron suppression for a given suppression in cps units—our WEH concentration modeling indicates that CSETN is less sensitive to hydrogen than the Sanin et al. (2017) modeling—so after the neutron suppression to WEH conversions, the two models yield nearly the same WEH concentration.

For instance, assume a single detector system derived with the Sanin et al. and Su modelings. Then the CL cps = [0.25, 0.233]. After the GCR and UL background subtraction we assume a neutron suppression of 0.05 cps. Their respective neutron suppression is $(0.25 - 0.05)/0.25$, and $(0.233 - 0.05)/0.233 = [0.80, 0.785]$. Conversion of neutron suppression to WEH concentration uses the Sanin conversion parameters [$a, b, c$] = [1.2, 0.06, $-0.51$]. The Su et al. parameters are [$a, b, c$] = [1.292, 0.0977, $-0.54$]. After the conversions to WEH concentration—we get 0.447 wt% versus 0.424 wt%. The difference in the two WEH concentrations is thereby negligible and justifies our use of the Sanin et al. parameters and conversions used in this paper.

It is important to state that the interpretation of WEH at a given location is highly model dependent. Layering, the form, and mixture of a given deposit with regolith may influence the interpretation. Regolith geochemistry is assumed to be constant and uniformly distributed within the detectable top meter of regolith. The assumption is potentially an important factor for the anhydrous region definition and for the differential method for determining WEH concentration. The differential method assumes that the only source of variation between the two contrasted surfaces is due to hydrogen variation in the regolith; in this case, the 65–70° S latitude band over which the averaged count rate is determined, Equation (2). The band subtends both feldspathic highlands terrain and the mafic South Pole Aitken basin. The two regions have well-known epithermal neutron counting rate differences (Lawrence et al. 2022). Their averaged composition and counting rates may not be consistent with the polar geochemistry where this paper's polar PSR analysis was performed.

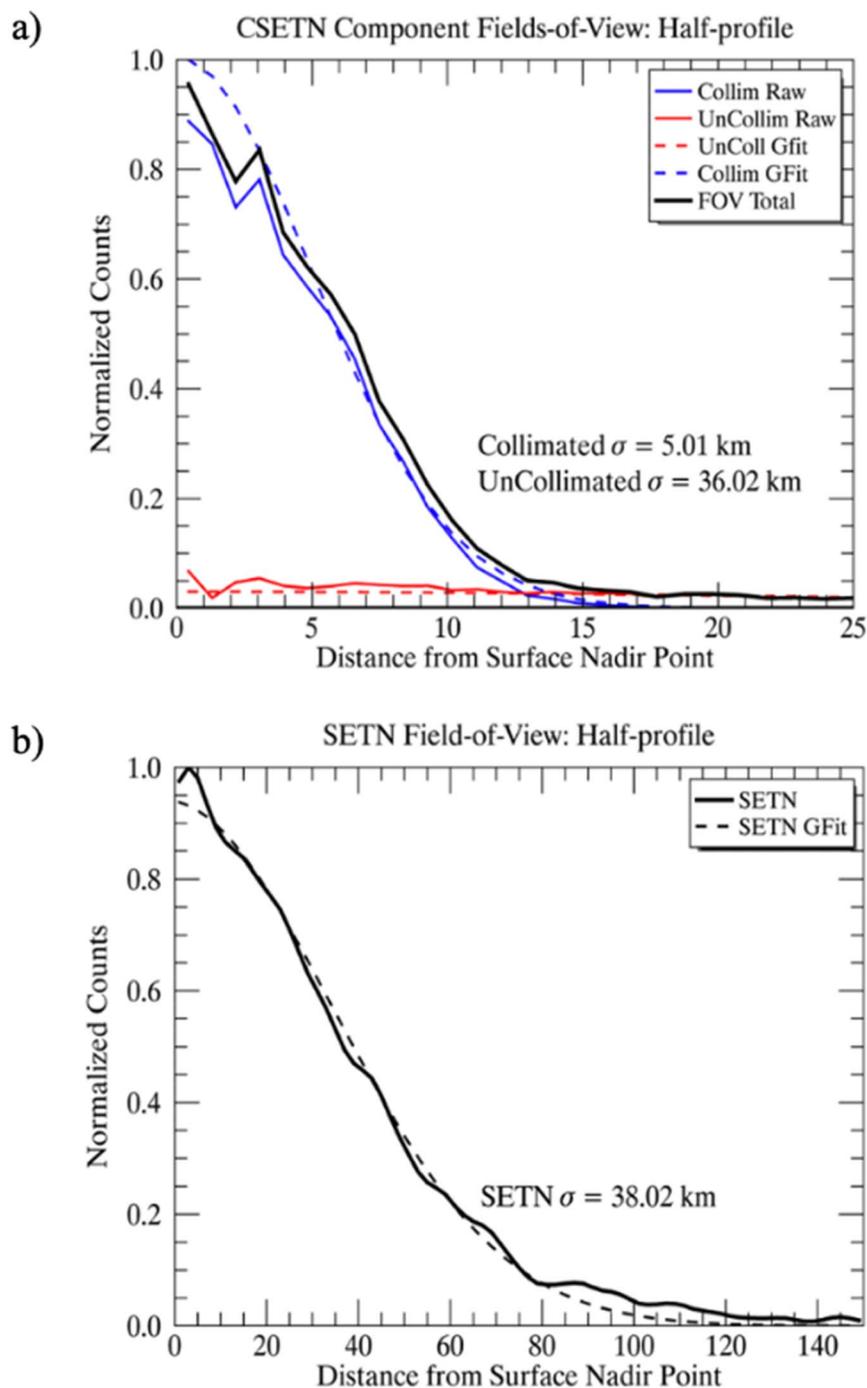

**Figure A7.** Half-profiles of the CSETN and SETN detectors' FOVs derived from our GEANT4 LEND modeling. Plots show CSETN's and SETN's detection of the lunar surface neutron-emission flux, as measured from a 50 km altitude and nadir pointing (x-axis = 0) over Ferroan ANorthositic (FAN) regolith. (a) The CSETN modeled total spatial response black is a convolved mixture (blue, red) of collimated and uncollimated neutrons. CSETN's collimated neutrons are sourced from surfaces less than 15 km from nadir, with FWHM = 11.8 km. Small profile deviations observed from the CSETN Gaussian fits (dashed) are due to the CSETN collimator solid geometry variation on the detection of lunar neutrons. The plots show the increasing distance to the lunar surface emission point away from nadir, moving right along the plot. (b) SETN is an uncollimated sensor. SETN has a broad FOV, with FWHM = 89.3 km. Note that the provided software, CSETN_FOV_Kern.pro, derives the uncollimated kernel shown in Figure 2 using the CSETN plot distributions.

Regolith properties, including densities, grain sizes, and porosities, are assumed to be constant for all regions being analyzed. Hydrogen concentrations are assumed to be uniformly distributed within the top meter of the regolith. Hydrogen layering and the depth of the enhanced-WEH layer

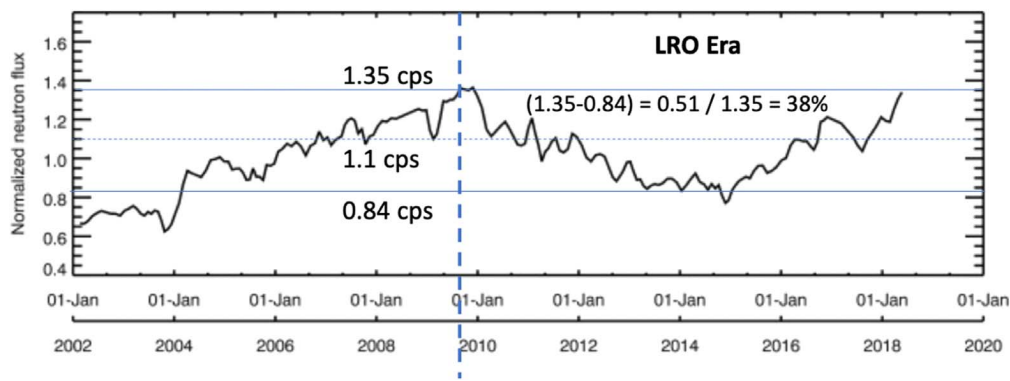

**Figure A8.** Normalized neutron flux at Mars as observed by the High Energy Neutron Detector. Source figure is annotated from Litvak et al. (2020).

(s) may cause ambiguous interpretations. For instance, relative neutron enhancement may be observed if there is a top layer of water ice. Regolith properties may also vary between slopes where rocky materials may be located or be more fine grained in the crater basins. Surfaces with more meteorics exposure have been shown to have smaller grains sizes (O'Brien & Byrne 2022). The variability may imply different regolith densities or textures, which may influence their respective neutron-emission flux. Further, variations in the GEANT4 modeling factors, neutron-scattering cross sections, instrument geometries and compositions, spacecraft materials or compositions, or other dependent parameters can produce important model differences.

Lastly, in this study we found excellent agreement between CSETN's largest PSR observations, when corrected with the linear model of Figure 8, and the LPNS restored observations (Elphic et al. 2007; Eke et al. 2015), e.g., Cabeus-1 and the Shoemaker PSR, which is the largest and most detectable PSRs by CSETN. If we assume that CSETN's collimated count rate was significantly lower than what is presented in the paper, as proposed in criticisms of CSETN's performance (Lawrence et al. 2010, 2022; Eke et al. 2012; Miller 2012; Teodoro et al. 2014), then the relative collimated neutron suppression would be greater, which would increase their WEH concentration (Sanin et al. 2017).

*Systematic sources of error.* We assume CSETN's ground detector calibrations that are ongoing at the University of Arizona, Lunar and Planetary Laboratory as directed by LEND co-PI, William Boynton, accurately correct for the evolving and variant combinations of observational conditions experienced during LEND's now well over a decade-spanning south pole observational campaign.

As shown in Figures A2 and A3, LRO's orbital inclination, operating altitude, and operating longitudes have evolved substantially during the mission. We know that the signal-to-noise ratios of the CSETN detectors are different due in part to their location relative to the spacecraft geometry and dependent production of neutrons and charged particles. Figure A8 shows that nearly a complete solar cycle occurred between the LRO start of operations in 2009 July approaching 2019. The period shows that an expected change of 38% change in the Mars neutron-emission flux occurred between solar minima at the end of 2009 and solar maximum near 2015. A similar change in the neutron-emimsson flux would be expected for the inner solar systems' airless small bodies.

Boynton et al. (2012) describes an exponential fitting approach to correct the $^3$He detector's efficiency recovery after LEND power cycling. The fits are performed for each detector on polar orbital averages that span the efficiency recovery period, generally 2 weeks (early mission), between LEND power cycles. Each orbital average is taken over a range of latitudes over the south pole. LRO's orbits are 2 hr in duration, which is the sampling time resolution for the exponential fits. GCR variation can vary by several percent, at timescales that are on the order of minutes, which is less than the orbital averages used for calibration. Thereby, the fits may not be fully responsive to the higher time resolution GCR variation. This source of error is expected to average out over the long history of the mission.

Systematically varying observing conditions are a potential source of error. We assume that ground calibration processing fully corrects CSETN's detector observations for regions that were sampled under very differing observing conditions. A case in point is that the top 3°–4° of south pole latitudes were highly observed in the first 2 years with up to four detectors. Plus, the first several months of observations 2009 July to 2009 December were taken during the extended solar minimum, where the lunar neutron count rates were highest (see the figure above). During that time LRO station keeping near 90° E and 270° E longitudes eliminated the coverage around those longitudes. Power cycling then strongly reduced the early mission coverage of Faustini's PSR. As a result, Faustini's PSR was likely not strongly observed until several years into the mission (Figure 5(d)).

After 2011, only two of CSETN's detectors were in operation, the orbital inclination began to degrade, and the top few degrees of latitude and the largest PSRs got progressively lesser coverage and were no longer observed after 2013. So, for areas toward equatorial latitudes below 86° S, the mapping campaign fully spans the decade. Planning ephemeris indicates ongoing coverage of the region equatorward of 85° S but at mostly a higher altitude > 50 km.

LRO delta-PSRs maneuvers were randomized after 2011 as to the longitude in which they were performed. This change enabled the subsequent and ongoing observation campaign of the 90 and 270° E longitudes.

### ORCID iDs

T. P. McClanahan  https://orcid.org/0000-0003-3708-4028